\documentclass[twoside,english]{elsarticle}
\usepackage[T1]{fontenc}
\usepackage[latin9]{inputenc}
\usepackage{geometry}
\usepackage{array}
\usepackage{amsmath}
\usepackage{graphicx}
\usepackage{setspace}
\makeatletter

\providecommand{\tabularnewline}{\\}

\usepackage{babel}

\makeatletter
\def\ps@pprintTitle{%
\let\@oddhead\@empty
\let\@evenhead\@empty
\def\@oddfoot{}%
\let\@evenfoot\@oddfoot}
\makeatother
\makeatletter

\@ifundefined{showcaptionsetup}{}{%
 \PassOptionsToPackage{caption=false}{subfig}}
\usepackage{subfig}
\makeatother

\usepackage{babel}
\begin{document}
\begin{frontmatter}{}

\title{Pattern Formation in the Longevity-Related Expression of Heat Shock
Protein-16.2 in \emph{Caenorhabditis elegan}s}

\author[appmcu,iqbiocu]{J.~M. Wentz}

\author[fhcrc]{A. Mendenhall}

\author[appmcu]{D. M.~Bortz\corref{cor1}}

\cortext[cor1]{Corresponding author: \emph{dmbortz@colorado.edu}}

\address[appmcu]{Department of Applied Mathematics, University of Colorado, Boulder,
CO 80309-0526, USA}

\address[iqbiocu]{Interdisciplinary Quantitative Biology Graduate Program, University
of Colorado, Boulder, CO 80309-0596, USA}

\address[pathuw]{Department of Pathology, University of Washington, Seattle, WA 98109-1024,
USA}
\begin{abstract}
Aging in \emph{Caenorhabditis elegans }is controlled, in part, by
the insulin-like signaling and heat shock response pathways. Following
thermal stress, expression levels of small heat shock protein 16.2
show a spatial patterning across the 20 intestinal cells that reside
along the length of the worm. Here, we present a hypothesized mechanism
that could lead to this patterned response and develop a mathematical
model of this system to test our hypothesis. We propose that the patterned
expression of heat shock protein is caused by a diffusion-driven instability
within the pseudocoelom, or fluid-filled cavity, that borders the
intestinal cells in \emph{C. elegans. }This instability is due to
the interactions between two classes of insulin like peptides that
serve antagonistic roles.\emph{ }We examine output from the developed
model and compare it to experimental data on heat shock protein expression.
Furthermore, we use the model to gain insight on possible biological
parameters in the system. The model presented is capable of producing
patterns similar to what is observed experimentally and provides a
first step in mathematically modeling aging-related mechanisms in
\emph{C. elegans.}
\end{abstract}
\end{frontmatter}{}

\section{Introduction}

Aging in \emph{Caenorhabditis elegans} is linked to many physiological
processes including the stress response and insulin-like signaling.
The expression levels of heat shock proteins (HSP) in \emph{C. elegans}
decrease with age, suggesting HSP expression is a biomarker of aging
\citep{Lund2002}. In isogenic populations of \emph{C. elegans} a
fourfold variation in lifespan following thermal stress is predicted
by the intestinal transcription of gene that encodes a small heat
shock protein, HSP-16.2 \citep{Rea2005}. The expression of \emph{hsp-16.2}
is not uniform across all intestinal cells but rather shows distinct
spatial patterning, with magnitudes that vary in conjunction with
the worm's lifespan \citep{Seewald2010}. Patterns consistently show
elevated expression of \emph{hsp-16.2} in the anterior and posterior
intestinal cells and decreased expression in the middle intestinal
cells \citep{Mendenhall2015}. The mechanism behind the patterned
expression is currently unknown. With the use of mathematical modeling,
we aim to explore possible mechanisms that lead to the emergence of
\emph{hsp-16.2} patterns in intestinal cells.

The heat shock response is characterized by the increased production
of HSP. Many HSP act as chaperones and help stabilize and refold proteins
damaged by thermal pressure and other stressors \citep{Haslbeck2005}.
The heat shock response is activated both autonomously within a cell
and via intercellular signaling pathways. Recent evidence suggests
that, in \emph{C. elegans} subjected to heat shock, neuronal signals
override the cell autonomous heat shock response \citep{Prahlad2011,Prahlad2008a}.
Furthermore, stimulation of thermosensory and serotonergic neurons
is sufficient for inducing the heat shock response in peripheral cells
in the absence of temperature changes \citep*{Tatum2015}. The neuronal
control of the heat shock response is dependent on the release of
serotonin from the ADF chemosensory neurons and neurosecretary motor
(NSM) neurons \citep{Tatum2015}. This evidence suggests that the
\emph{hsp-16.2} production observed following thermal injury is the
result of neuronal cues that lead to the release of serotonin. The
released serotonin then sends signals to distal cells to trigger the
heat shock response. In this paper we are interested in exploring
the mechanism by which serotonin released at the head of the worm
can activate the patterned expression of heat shock proteins in intestinal
cells microns away from the head.

We hypothesize that the neuronal regulation of the heat shock response
acts through the insulin-like signaling pathway \citep{Prahlad2009}.
Studies have shown that the serotonin released by the ADF and NSM
neurons indirectly stimulates the nuclear translocation of DAF-16
in distal cells through the insulin-like DAF-2 receptor \citep{Liang2006}.
Furthermore, DAF-16 likely plays a role in the transcription of small
HSP genes, including \emph{hsp-16.2} \citep{Hsu2003,Hartwig2009,Walker2003}.
This mechanism is further complicated by the presence of 40 insulin
like peptides (ILPs) in \emph{C. elegans} that may serve as either
DAF-2 agonists or antagonists. The activation of DAF-2 results in
blocking DAF-16 nuclear localization. In turn, decreasing nuclear
DAF-16 has a direct impact on the level of ILP transcription \citep{Tepper2013}.
These interactions suggest that ILPs regulate their own production
through positive and/or negative feedback loops.

We propose that the spatial patterning of \emph{hsp-16.2} transcription
along intestinal cells arises due to a diffusion-driven instability.
Diffusion-driven or Turing instabilities provide an explanation for
how patterns can emerge in biological systems \citep{Murray2000}.
Researchers have used these instabilities to explain a wide range
of biological phenomenon, from embryonic development to animal coat
pattern formation. One class of reaction diffusion systems that is
known to produce patterns involves both an activator and an inhibitor
species. The activator stimulates its own production as well as the
production of the inhibitor, while the inhibitor blocks both its own
production and the production of the activator. Furthermore, the inhibitor
diffuses at a faster rate than the activator. We hypothesize that
the insulin-like signaling pathway falls into this class of reaction
diffusion systems. The set of ILPs that act as DAF-2 antagonists represents
the activator and the set of ILPs that act as DAF-2 agonists represents
the inhibitor. The interaction between the two classes of ILPs in
the presence of their diffusive properties leads to an instability
that results in patterned ILP concentrations. As a result DAF-2 is
activated in a spatially dependent manner. The transcription of \emph{hsp-16.2},
in turn, demonstrates a similar spatial patterning. 

Here, we develop and analyze a mathematical model to examine this
system, and we show that the model is capable of producing patterns
reminiscent of what is observed experimentally. In the methods, we
describe the mathematical model development and detail a new analysis
of previously collected data on \emph{hsp-16.2} transcription in \emph{C.
elegans}. We then present the results of the mathematical model and
examine the required parameter constraints to obtain diffusion driven
instability. We compare the model predictions to the analyzed data
and conclude that the model successfully reproduces the \emph{hsp-16.2}
transcription patterns. Our results suggest that two groups of worms
exist with different \emph{hsp-16.2} expression profiles. This is
the first mathematical model of the neuronally controlled heat shock
response in \emph{C. elegans}. The model can be used to predict how
biological parameters affect HSP expression and, ultimately, can help
assess the effect of specific biological perturbations on aging. 

\section{Methods\label{sec:Methods}}

\subsection{Model Development\label{subsec:Model-Development}}

The standard framework for the study of pattern formation in reaction-diffusion
equations involves an activator $A$ and an inhibitor $B$. For a
system with two spatial dimensions, this can be qualitatively described
by 
\begin{eqnarray}
A_{t}(t,x,y) & = & \text{activator diffusion + activator reactions}\nonumber \\
B_{t}(t,x,y) & = & \text{inhibitor diffusion + inhibitor reactions}\label{eq:outline}
\end{eqnarray}
where the subscripted $t$ represents the derivatives of these concentrations
with respect to time. We can more quantitatively describe the model
using a general conservation law model: 
\begin{eqnarray}
A_{t}(t,x,y) & = & D_{A}\nabla^{2}A+F(A,B)\nonumber \\
B_{t}(t,x,y) & = & D_{B}\nabla^{2}B+G(A,B)\,,\label{eq:outline_quant}
\end{eqnarray}
where $D_{A}$ and $D_{B}$ are diffusion coefficients and $F$ and
$G$ are functions modeling the activator and inhibitor reactions,
respectively.

\begin{figure}
\begin{centering}
\includegraphics[bb=0bp 300bp 500bp 540bp,clip,width=0.6\columnwidth]{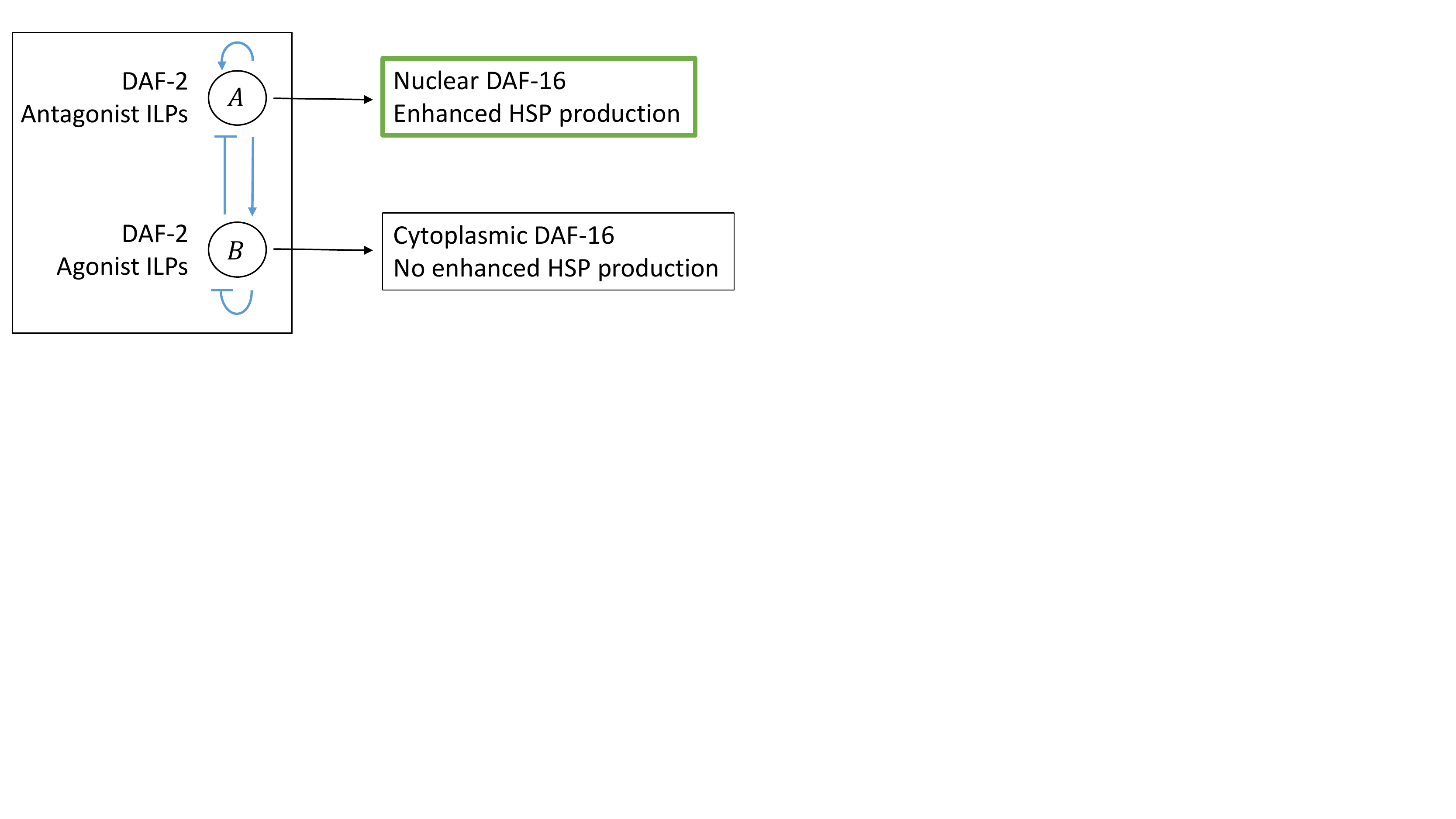}
\par\end{centering}
\caption{Interactions between the activator class $A$ and inhibitor class
$B$ in the reaction diffusion model. The activator $A$ leads to
the production of HSPs while the inhibitor $B$ does not.}
\label{fig:molecularDiagram}
\end{figure}

In this framework, the activator $A$ represents the ILPs that act
as DAF-2 antagonists, and the inhibitor $B$ represents ILPs that
act as DAF-2 agonists (Figure \ref{fig:molecularDiagram}). The DAF-2
antagonists activate the production of both classes of ILPs by allowing
DAF-16 nuclear translocation. The DAF-2 agonists block DAF-16 nuclear
translocation, and, thus, inhibit the production of both classes of
ILPs. The \emph{C. elegans} genome contains 40 genes encoding possible
ILPs. Of these genes, eight have been shown to be transcriptionally
regulated by DAF-16 \citep{Tepper2013,Murphy2003a,Kaletsky2015}.
For example, both \emph{ins-6} and \emph{ins-18} are positively regulated
by DAF-16. INS-6 acts as a DAF-2 agonist \citep{Hua2003}, and is
therefore a member of the inhibitor class $B$, and INS-18, a DAF-2
antagonist \citep{Pierce2001}, belongs to the activator class $A$.
The only ILP gene that is known to decrease in response to DAF-16
nuclear translocation is \emph{ins-7}, which codes a DAF-2 agonist.
Thus, the characteristics of INS-7 do not clearly match the characteristics
of the inhibitor class. However, we hypothesize that, although \emph{ins-7}
is downregulated by nuclear DAF-16, the net effect of DAF-16 on ILPs
that are DAF-2 agonists is stimulatory. An alternative model was explored
in which DAF-16 led to a net decrease in DAF-2 agonist production,
but we found that this model was not capable of pattern formation
(See \ref{sec:AppAltMechanism}). 

To mathematically model the reactions in the system, we first derive
equations for the extracellular DAF-2 and ILP interactions and then
incorporate the intracellular effects when DAF-2 is activated. We
use Michaelis-Menten kinetics with competitive inhibition to model
the extracellular binding of the two classes of ILPs to the DAF-2
receptor. We assume the binding of ILPs to DAF-2 happens on a much
faster time scale than other reactions in the system. In this analysis
we use the amount of agonist ($B$) that is bound to DAF-2, denoted
as $BR$, as a proxy for the activation of an intracellular pathway.
The concentration of ILP agonist bound to the DAF-2 receptor is approximated
by the following Michaelis-Menten relationship,
\begin{equation}
BR=\frac{R_{0}B}{K_{D,B}(1+\frac{A}{K_{D,A}})+B},\label{eq:BR}
\end{equation}
where $K_{D,B}$ is the dissociation constant of the DAF-2 agonist
class of ILPs, $K_{D,A}$ is the dissociation constant of the DAF-2
antagonist ILPs, and $R_{0}$ is the total concentration of DAF-2
receptors. To relate Equation \ref{eq:BR} to an intracellular pathway,
we introduce a new variable $P$ that represents an intracellular
product formed by the activation of DAF-2. The change in $P$ with
time is approximated as
\begin{equation}
\frac{dP}{dt}=\frac{V_{max}B}{K_{D,B}(1+\frac{A}{K_{D,A}})+B}-k_{7}P,\label{eq:dPdt}
\end{equation}
where $V_{max}$ is the maximum rate of product formation, and $k_{7}$
is the specific rate at which the product $P$ leaves the system.
We assume the system rapidly reaches equilibrium and, thus, calculate
$P$ explicitly by setting Equation \ref{eq:dPdt} equal to zero
\begin{equation}
P=\frac{V_{max}B}{k_{7}(K_{D,B}(1+\frac{A}{K_{D,A}})+B)}.\label{eq:P}
\end{equation}

When DAF-2 is activated, a complex set of intracellular processes
takes place that leads to the phosphorylation of DAF-16, blocking
nuclear translocation \citep{Ewbank2006}. We note that the product
$P$ can be viewed as an intermediate molecule in this process such
as, for example, AGE-1. For the mathematical model, we simplify this
system and incorporate a direct effect of $P$ on the production of
$A$ and $B$. As the amount of $P$ in the system increases there
is less nuclear DAF-16 (assuming total DAF-16 within a cell is constant)
and therefore the production of $A$ and $B$ decreases. We use a
hill function to describe this effect and write the final reactions
as
\begin{align}
F(A,B) & =k_{1}-k_{2}\frac{P^{r}}{H^{r}+P^{r}}-k_{3}A\label{eq:FAB}\\
G(A,B) & =k_{4}-k_{5}\frac{P^{r}}{H^{r}+P^{r}}-k_{6}B.\label{eq:GAB}
\end{align}
where $H$ is the concentration of $P$ at which the effect of $P$
on the nuclear translocation of DAF-16 is half maximized, and $r$
is the hill-coefficient that determines the steepness of the hill
function.

\begin{figure}
\begin{centering}
\includegraphics[bb=0bp 100bp 960bp 540bp,scale=0.3]{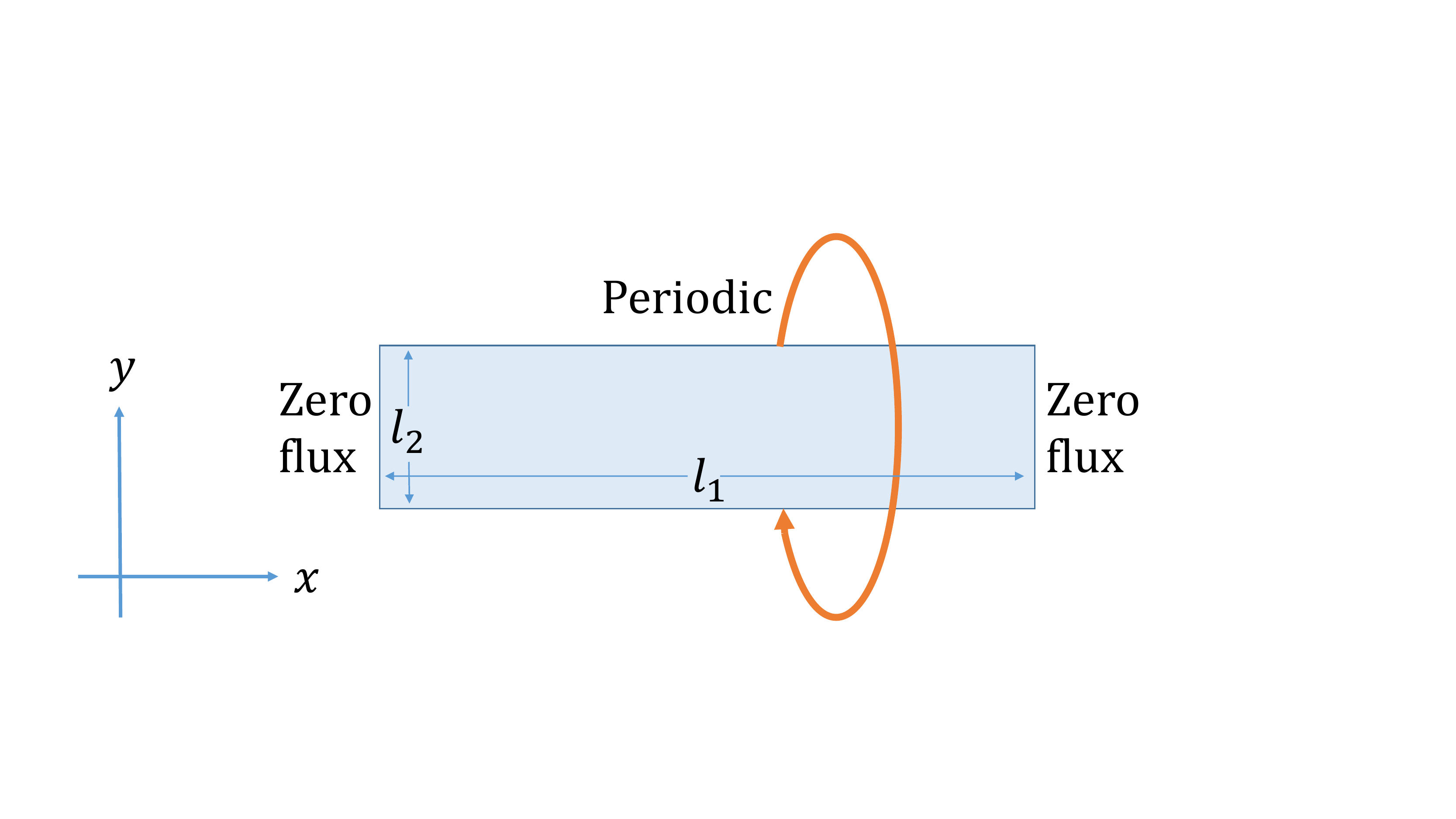}
\par\end{centering}
\caption{Domain on which the reaction diffusion model was solved with boundary
conditions shown.\label{fig:domain}}
\end{figure}

The system given by Equations \ref{eq:outline_quant}, \ref{eq:P},
\ref{eq:FAB}, and \ref{eq:GAB} was made dimensionless using the
following substitutions, 
\begin{align*}
t^{*} & =D_{A}t/L^{2} & x^{*} & =x/L & y^{*} & =y/L\\
d & =D_{B}/D_{A} & \gamma & =k_{3}L^{2}/D_{A} & u & =Ak_{3}/k_{1}\\
v & =Bk_{3}/k_{4} & p & =Pk_{7}/V_{max} & a & =k_{2}/k_{1}\\
b & =k_{5}/k_{4} & c & =k_{6}/k_{3} & h & =H(k_{7}/V_{max})\\
k_{D,A} & =K_{D,A}k_{3}/k_{1} & k_{D,B} & =K_{D,B}k_{3}/k_{4},
\end{align*}
resulting in the following set of equations 
\begin{align}
\frac{\partial u}{\partial t^{*}}=\gamma f(u,v)+\nabla^{2}u\label{eq:dudt}\\
\frac{\partial v}{\partial t^{*}}=\gamma g(u,v)+d\nabla^{2}u\label{eq:dvdt}
\end{align}
where 
\begin{align}
f(u,v) & =1-a\frac{p(u,v)^{r}}{h^{r}+p(u,v)^{r}}-u\label{eq:fuv}\\
g(u,v) & =1-b\frac{p(u,v)^{r}}{h^{r}+p(u,v)^{r}}-cv\label{eq:guv}\\
p(u,v) & =\frac{v}{k_{D,B}(1+\frac{u}{k_{D,A}})+v}.\label{eq:puv}
\end{align}
Here $L$ represents the length of the worm. For notational simplicity,
from here forward, we will use $x$ and $y$ to represent $x^{*}$
and $y^{*}$, respectively. We determined approximate solutions to
the PDE by linearizing the system about the spatially homogeneous
steady state $(u_{0},v_{0})$, calculated by setting Equations \ref{eq:fuv}
and \ref{eq:guv} equal to zero. The steady state of this system was
calculated numerically using a Newton method as implemented in the
nlqslv package of R \citep{Hasselman2016}. We solved the linearized
system on a cylindrical domain (Figure \ref{fig:domain}). 
The shape of the domain was approximated based on the dimensions of
\emph{C. elegans}, where the length of the domain $l_{1}$ represents
the length of the worm and the height of the domain $l_{2}$ represents
the circumference of the worm. Since we are working in nondimensional
coordinates where the characteristic length $L$ is equal to the length
of the worm, $l_{1}=1.$ In \emph{C. elegans} the length to maximum
diameter ratio is approximately 18 \citep{Hirose2003}. Thus we set
$l_{2}=0.2l_{1}$. The following boundary conditions are imposed on
the domain (\textbf{$\mathbf{W}=[u-u_{0},v-v_{0}]$}):
\begin{align*}
\frac{d\mathbf{W}}{dx}_{x=0,l_{1}} & =0\\
\frac{d\mathbf{W}}{dy}_{y=0} & =-\frac{d\mathbf{W}}{dy}_{y=l_{2}}.
\end{align*}
Periodic boundary conditions in the $y$ direction are used to approximate
diffusion on a 2D cylindrical surface. Using these boundary conditions,
the solutions to the linear problem take the form 
\begin{equation}
\mathbf{w}(x,y,t)=\sum_{k}c_{k}e^{\lambda t}\mathbf{W}_{k}(x,y)
\end{equation}
where the time independent eigenfunctions are 
\begin{equation}
\textbf{W}_{k}(x,y)=\textbf{B}\cos\left(\frac{n\pi}{l_{1}}x\right)\left(\textbf{C}\sin\left(\frac{2m\pi}{l_{2}}y\right)+\textbf{D}\cos\left(\frac{2m\pi}{l_{2}}y\right)\right)\label{eq:eigenfunctions}
\end{equation}
and the wavenumber $k$ is given by 
\begin{equation}
k^{2}=\left(\frac{n\pi}{p}\right)^{2}+\left(\frac{2m\pi}{q}\right)^{2}.\label{eq:wavenumber}
\end{equation}
Here $\lambda$ is the eigenvalue that determines the temporal growth
of the modes and it is given by the roots of 
\begin{equation}
\lambda^{2}+\lambda(k^{2}(1+d)-\gamma(f_{u}+g_{v}))+dk^{4}-\gamma(df_{u}-g_{v})k^{2}+\gamma^{2}(f_{u}g_{v}-f_{v}g_{u})=0\label{eq:dispersionRelation}
\end{equation}
(see \citep{Murray2000} for details). The partial derivatives of
$f$ and $g$ are calculated at the steady state values of the system
$(u_{0},v_{0})$. The expression $\lambda(k^{2})$ will be referred
to as the \emph{dispersion relation}. Modes with $\lambda(k^{2})>0$
grow as time increases, and represent the possible patterns that can
be produced.

We used the CountourPlot3D and RegionPlot3D functions in Mathematica
to numerically identify regions of parameter space that can lead to
pattern formation \citep{WolframResearch2016}. These functions initially
evaluate the criteria for pattern formation on a 3D grid and then
uses an adaptive algorithm to evaluate the function on select subdivided
regions to obtain the regions where the specified criteria are met.
We generated two sets of region plots. First we examined which reaction
parameters satisfied the Routh-Horwitz criterion. Then we incorporated
spatial properties, including diffusion and domain size, to determine
how these properties affect the number of possible modes that can
result.

\subsection{Data Analysis}

We compared our model output to data on \emph{hsp-16.2} transcription
in \emph{C. elegans} \citep{Mendenhall2015}. This previous study
obtained relative \emph{hsp-16.2} transcription levels in each of
the 20 intestinal cells for 28 worms. Expression was measured 24 hours
after heat shock using a \emph{C. elegans }strain that contained a
\emph{hsp-16.2} promoter bound to a gene for green fluorescent-protein
(GFP). Thus, GFP fluorescence measured in live worms using microscopy
serves as a marker for \emph{hsp-16.2} transcription.

To analyze the available data, we first averaged the GFP signal in
cells within the same intestinal cell ring. There are a total of 9
intestinal cell rings in the \emph{C. elegans} intestine; 4 cells
in the anterior most cell ring and 2 cells in all the other cell rings.
Cells within the same cell ring are at approximately the same distance
down the length of the worm. Since our analysis predicts that variations
in HSP production are only observed along the length of the worm and
not along the circumference of the worm (see Section \ref{subsec:resultsModelAnalysis}),
we assume the rate of \emph{hsp-16.2} transcription is the same for
cells within a single ring. Next, we used a rolling window to average
across cell rings and smooth the data. For interior cell rings, the
window length was three. For example, the value at the 4th cell ring
was averaged with the 3rd and 5th to obtain an updated value at the
4th cell ring. For the edge cell rings we averaged the value of that
cell ring with its one neighbor. Finally we normalized the data for
each worm by dividing by the mean \emph{hsp-16.2} transcription value
across all intestinal cells. This analysis resulted in nine spatial
data points for each worm.

In order to compare this data to predictions from the mathematical
model, we decomposed the data into the relevant trigonometric modes.
Due to the zero flux boundary conditions of the system, we used only
cosine modes for the analysis (see Equation \ref{eq:eigenfunctions}).
Here, again we assume that $m=0$ and thus the system reduces to a
one dimensional problem. The nine normalized data points obtained
for each worm lead to the inclusion of nine modes in the projection
analysis. The contribution of each mode, $n=0,1,2,...,8$, is given
as
\begin{equation}
p_{n}(x)=\alpha_{n}\cos\left(\frac{n\pi(x-1)}{8}\right)
\end{equation}
where the expansion coefficients, $\alpha_{n},$ are
\begin{equation}
\alpha_{n}=\frac{1}{16}(y_{1}+\cos\left(n\pi\right)y_{9})+\frac{1}{8}\sum_{i=2}^{8}\cos\left(\frac{n\pi(i-1)}{8}\right)y_{i}\quad\quad n=0,8,
\end{equation}
\begin{equation}
\alpha_{n}=\frac{1}{8}(y_{1}+\cos\left(n\pi\right)y_{9})+\frac{1}{4}\sum_{i=2}^{8}\cos\left(\frac{n\pi(i-1)}{8}\right)y_{i},\quad\quad n=1,2,..,7.
\end{equation}
Here $y_{i}$ represents the normalized data from the $i$th intestinal
cell ring. Using the expansion coefficients obtained for each worm,
we grouped the worms using a k-means clustering analysis. We assessed
the validity of this clustering approach using a bootstrap analysis
as implemented in Flexible Procedures for Clustering package in R
\citep{Hennig2015}. 

\subsection{Model simulations}

To examine how patterns form and what modes dominate with time, we
simulated the reaction diffusion model using R v3.2.3 \citep{RCoreTeam2015b}.
The PDE was simulated over a 2D grid using a finite difference approach
as implemented in ReacTran (v1.3.2) \citep{Soetaert2009}. The grid
over which the system was solved was 100x20 cells. This resolution
corresponds to an approximate resolution of 10 $\mu m$ which is much
smaller than the length of an intestinal cell (approximately 100 $\mu m$)
but much larger than the size of a protein. To perform simulations
we used one of two initial conditions. In Case I the domain was initialized
using the homogeneous steady state values for each ILP class and Gaussian
noise was added. Specifically, each grid cell was initialized with
the following dimensionless quantity of the two classes of ILPs 
\begin{align*}
u_{ij} & =u_{0}+\epsilon_{u,ij}\\
v_{ij} & =v_{0}+\epsilon_{v,ij}
\end{align*}
where $\epsilon_{u,ij}$ and $\epsilon_{v,ij}$ were drawn from normal
probability distributions with means at $u_{0}$ and $v_{0}$ and
standard deviations of $0.10u_{0}$ and $0.10v_{0}$, respectively.
This level of variance was chosen to allow us to explore the effect
of noise on the evolution of the system when it is perturbed from
the homogeneous steady state. The initialization method avoids assumptions
about where the pattern is initialized. 

In Case II for model initialization, we explore our hypothesis that
the heat shock response is initialized by an ILP stimulus at the head
of the worm. Here, ILPs are added to one edge of the domain, corresponding
to a release at the head of the worm. In this case the first column
of the grid was initialized with the reaction steady-state values
and the rest of the grid had no ILPs. Specifically, if $u_{ij}$,
$v_{ij}$, represent the initial concentration of $u$, $v$, in grid
cell $(i,j)$, for i=1,2,...,20 the initialization concentrations
are given as 
\begin{align*}
u_{i1} & =u_{0}\\
v_{i1} & =v_{0}\\
u_{ij} & =0\quad\quad\quad j=2,3,...,100\\
v_{ij} & =0\quad\quad\quad j=2,3,...,100.
\end{align*}

Each simulation was run for for a dimensionless time $t^{*}$ of 2.0.
We chose this value of $t^{*}$ because it represents the same time
scale as the experimental data from \citep{Mendenhall2015}. Specifically,
we have that $t^{*}=tD_{A}/L^{2}$, and we estimate that the length
of the worm $L=1000$$\mu m$ and the diffusion rate $D_{A}$ is between
10 and 100 $\mu m^{2}/s$, i.e., the diffusion rate is between that
of a protein in water and that of a protein in a cell \citep{Milo2015}.
This corresponds to a time between 12 hours and 6 days, which is on
the same time scale as the data from the experiment which was obtained
24 hours post heat shock. When determining the steady-state of the
system, in some cases it was necessary to run the simulation for a
dimensionless time $t^{*}$of 3.0.

The source code for all the analyses perform is available on GitHub
\footnote{ReactionDiffusionCElegans repository from MathBioCU group on GitHub
(https://github.com/MathBioCU).}

\section{Results\label{sec:Results}}

\subsection{Reaction diffusion model predicts patterns along the length of the
worm\label{subsec:resultsModelAnalysis}}

The developed reaction diffusion model is capable of generating patterns
reminiscent of intestinal \emph{hsp-16.2} expression patterns in \emph{C.
elegans}. The model has a total of 9 dimensionless parameters (Table
\ref{tab:parameters}). 
\begin{table}
\begin{centering}
\caption{Parameters of the mathematical model. Values shown were used to perform
model simulations. The justification for these parameter values is
provided in the text. \label{tab:parameters}}
\par\end{centering}
\centering{}%
\begin{tabular}{|c|c|>{\raggedright}p{10cm}|}
\hline 
Parameter  & Value & \multicolumn{1}{>{\raggedright}p{10cm}}{Description}\tabularnewline
\hline 
\hline 
$a$  & 1 & Relative inhibitory effect of DAF-16 on activator based on maximum
production rate. The value implies activator production is completely
controlled by DAF-16.\tabularnewline
\hline 
$b$  & 1 & Relative inhibitory effect of DAF-16 on inhibitor based on maximum
production rate. The value implies inhibitor production is completely
controlled by DAF-16.\tabularnewline
\hline 
$c$  & 1 & Relative decay rate of inhibitor to activator. The value implies the
decay rate of the two classes of ILPs is the same.\tabularnewline
\hline 
$h$  & 0.4 & Concentration of $p$ at which the feedback strength is half maximized.
A low value implies feedback is very sensitive.\tabularnewline
\hline 
$r$  & 10 & Hill coefficient describing feedback strength. The large value implies
cooperativity exists in the system. \tabularnewline
\hline 
$k_{D,A}$  & 0.01 & Dimensionless dissociation constant for the activator.\tabularnewline
\hline 
$k_{D,B}$  & 0.01 & Dimensionless dissociation constant for the inhibitor.\tabularnewline
\hline 
$\gamma$  & 80 & Relative domain size. Reasonable biological values lie between 0.01
and 100 (see text).\tabularnewline
\hline 
$d$  & 5 & Ratio of diffusion rates between the two ILP classes. Implies the
inhibitor diffuses 5 times as fast as the activator\tabularnewline
\hline 
\end{tabular}
\end{table}
In order to explore the model's properties, where possible we set
parameter values equal to biologically reasonable values. The parameters
$a$ and $b$ were set equal to 1.0, implying that the expression
of ILPs is completely under the control of DAF-16. That is, as $p$
approaches infinity, all DAF-16 is blocked from entering the nucleus
and the net ILP production rate approaches zero. The parameter $c$
was set equal to one under the assumption that the two classes of
ILPs decay at the same rate (i.e., $k_{3}=k_{6}$). For the remaining
four parameters that describe the reactions in the system ($h$, $r$,
$k_{D,A}$, and $k_{D,B}$) we sought to find biologically reasonable
parameter values that together could result in diffusion driven instability. 

We determined the range of biologically feasible values for $k_{D,A}$
and $k_{D,B}$ by first determining ranges for the values of the dissociation
constants ($K_{D,A},$ $K_{D,B})$, ILP production rates ($k_{1},$
$k_{4}$), and ILP degradation rates ($k_{3}$) as follows. In humans,
insulin likely binds to its receptor in two stages, leading to two
dissociation constants of approximately 40 and 170 nM \citep{Subramanian2013}.
In our model we assume a single binding stage for all ILPs in \emph{C.
elegans} where the dissociation constant lies at a similar value (i.e.
between 40 and 170 nM). The degradation half-life of proteins can
vary drastically from 10 minutes to several days \citep{Belle2006,Cambridge2011}.
This leads to degradation rates that vary between $10^{-6}$ s$^{-1}$
and $10^{-3}$ s$^{-1}$. As for production rates, the maximum protein
production rate in \emph{E. coli} ranges from 1 to $10^{6}$ molecules
per generation, where the generation time is 21.5 minutes \citep{Li2014}.
To translate the production rate units from molecules to molarity,
we used a pseudocoelom volume range of 40 - 80 pL \citep{Banse2012}.
Putting this all together, we estimated that $k_{D,A}$ and $k_{D,B}$
values are likely between $10^{-3}$ and $10^{7}.$ This large range
is primarily due to the uncertainty in ILP production and degradation
rates. However, in subsequent sections, by using the mathematical
model and the requirements for pattern formation we were able to further
restrict this range.

Biological ranges for the values of $h$ and $r$ are more elusive.
The value of $h$ depends on $H,$ $V_{max},$ and the degradation
rate $k_{7}$. Both $H$ and $r$ are challenging to estimate since
they describe the effect of DAF-2 activation on DAF-16 nuclear translocation,
and the kinetics of this pathway have not been experimentally examined.
Thus, to explore the pattern formation space we set a lower bound
on $h$ of 0.05 but note that $h$ could potentially take on smaller
values. Furthermore, the largest value of $r$ we explore is 10, but
again this is not a biologically strict upper bound. 

\begin{figure}
\begin{centering}
\includegraphics[scale=0.6]{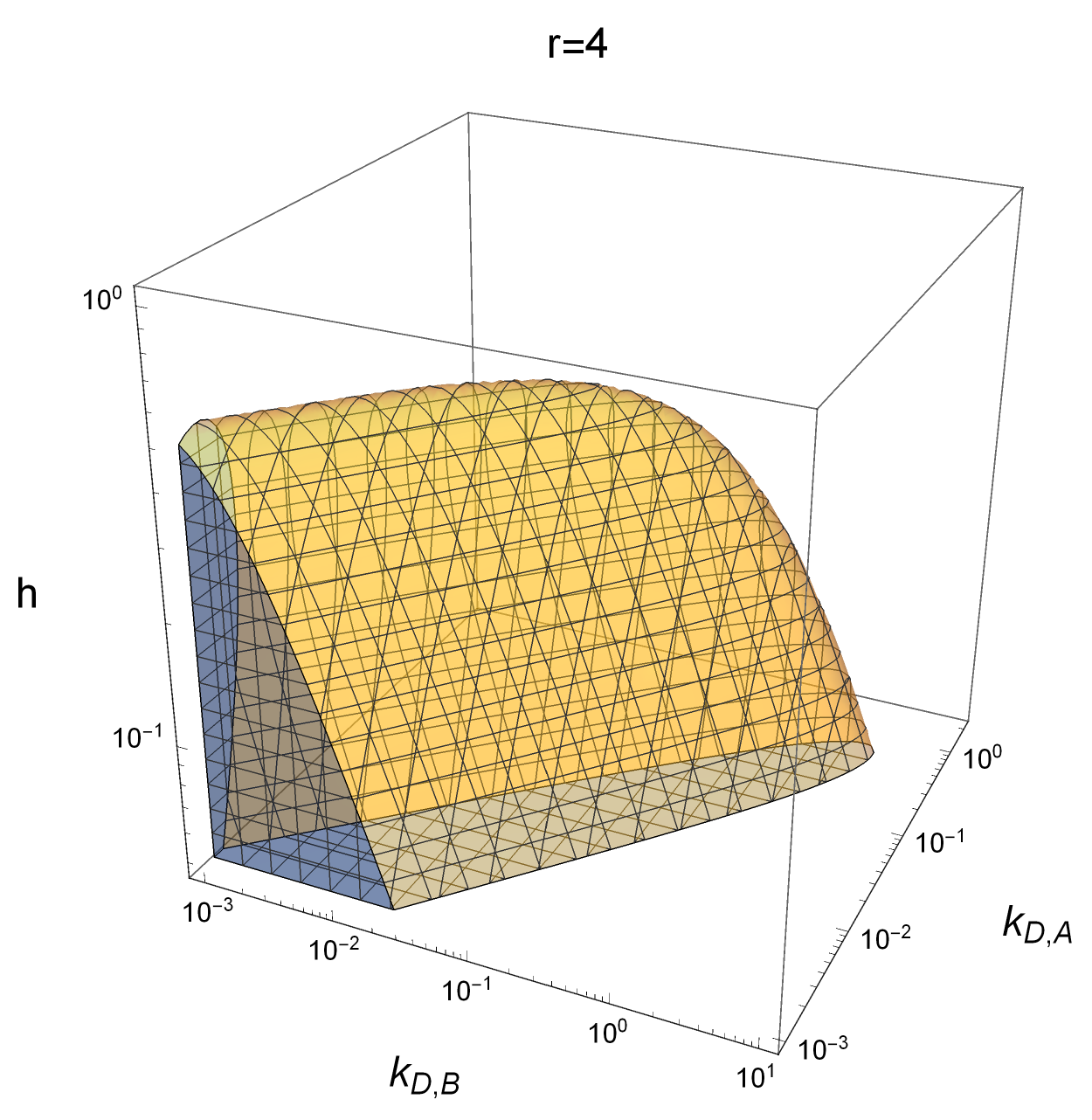}\includegraphics[scale=0.6]{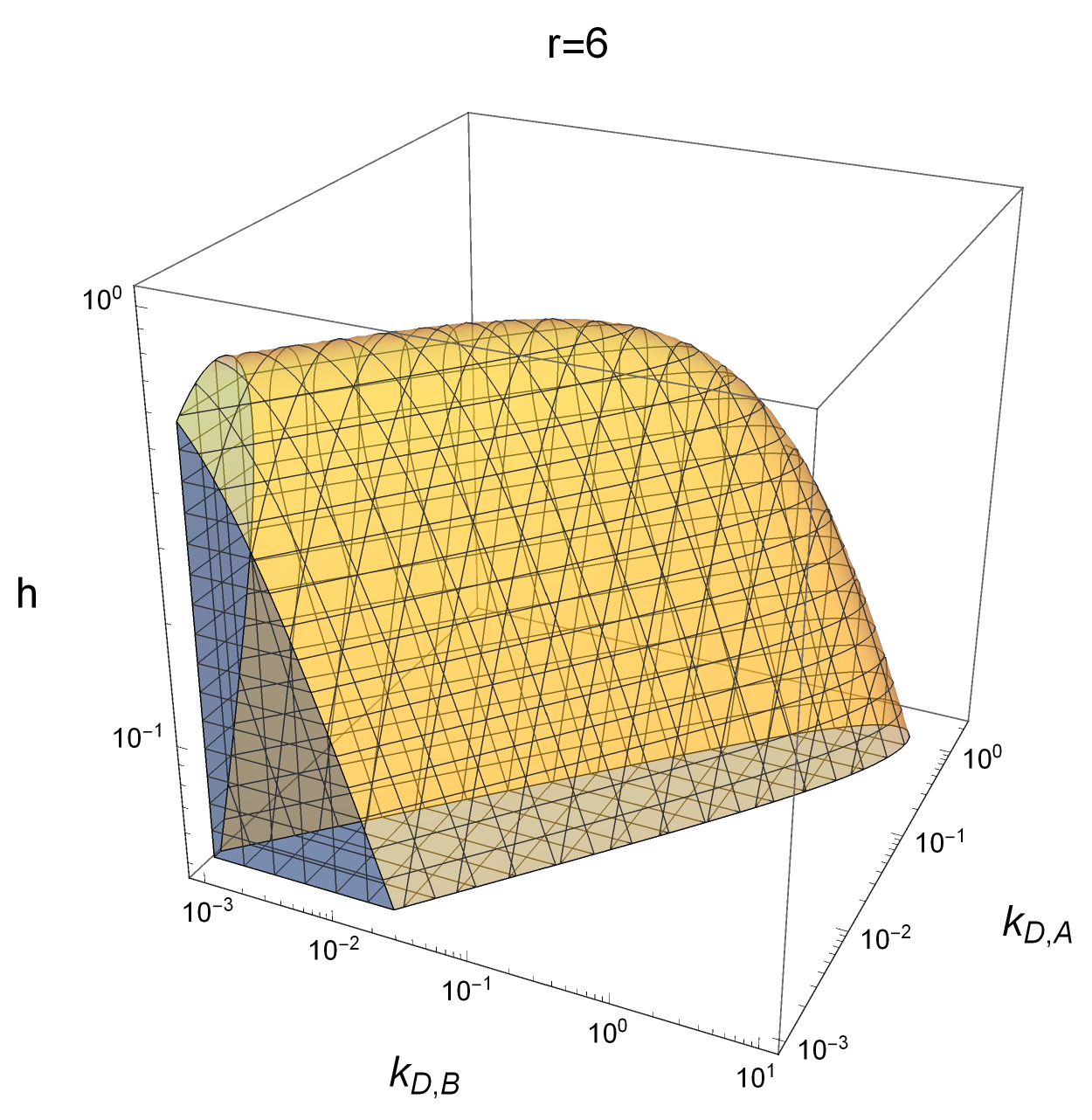}
\par\end{centering}
\begin{centering}
\includegraphics[scale=0.6]{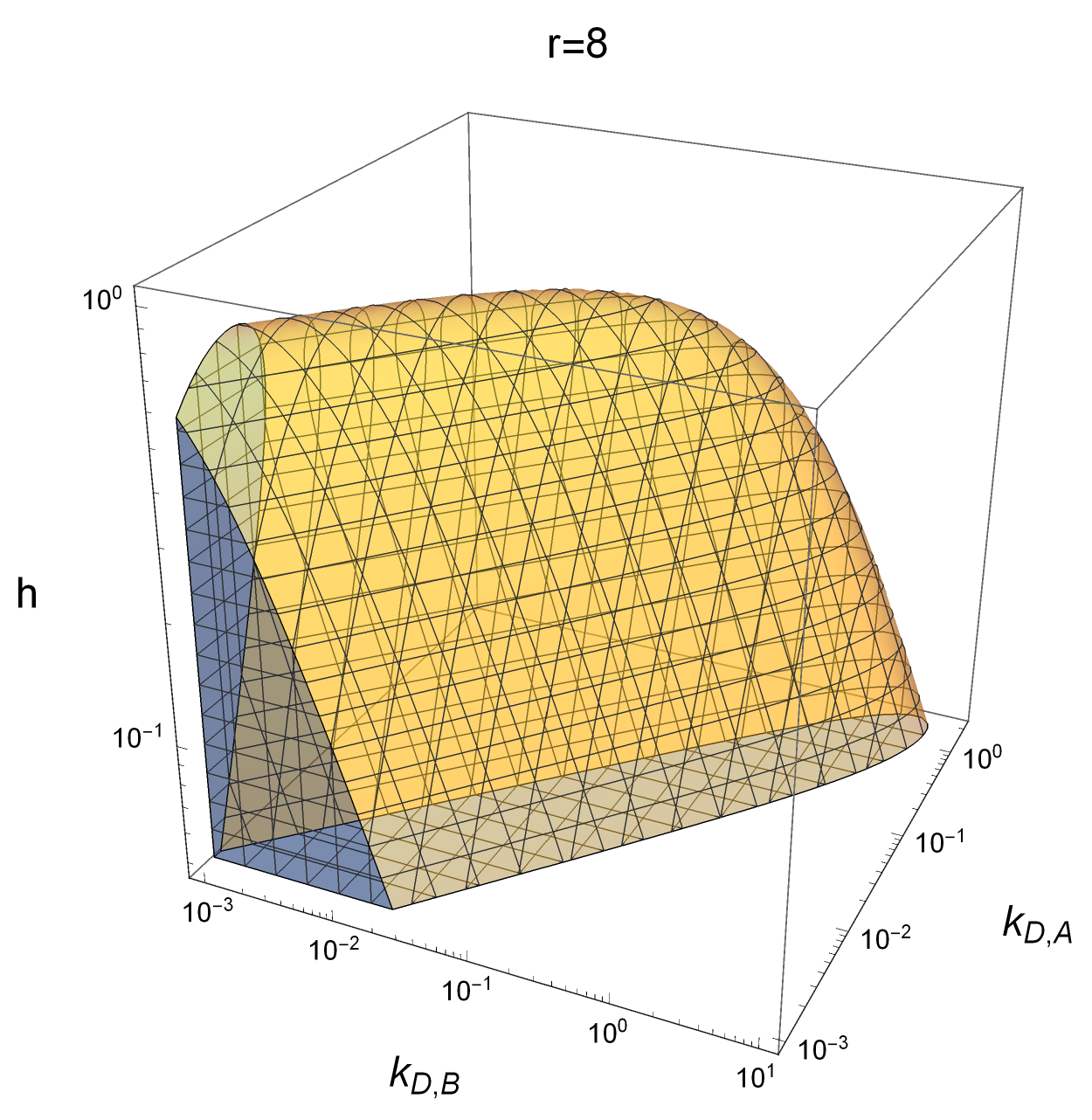}\includegraphics[scale=0.6]{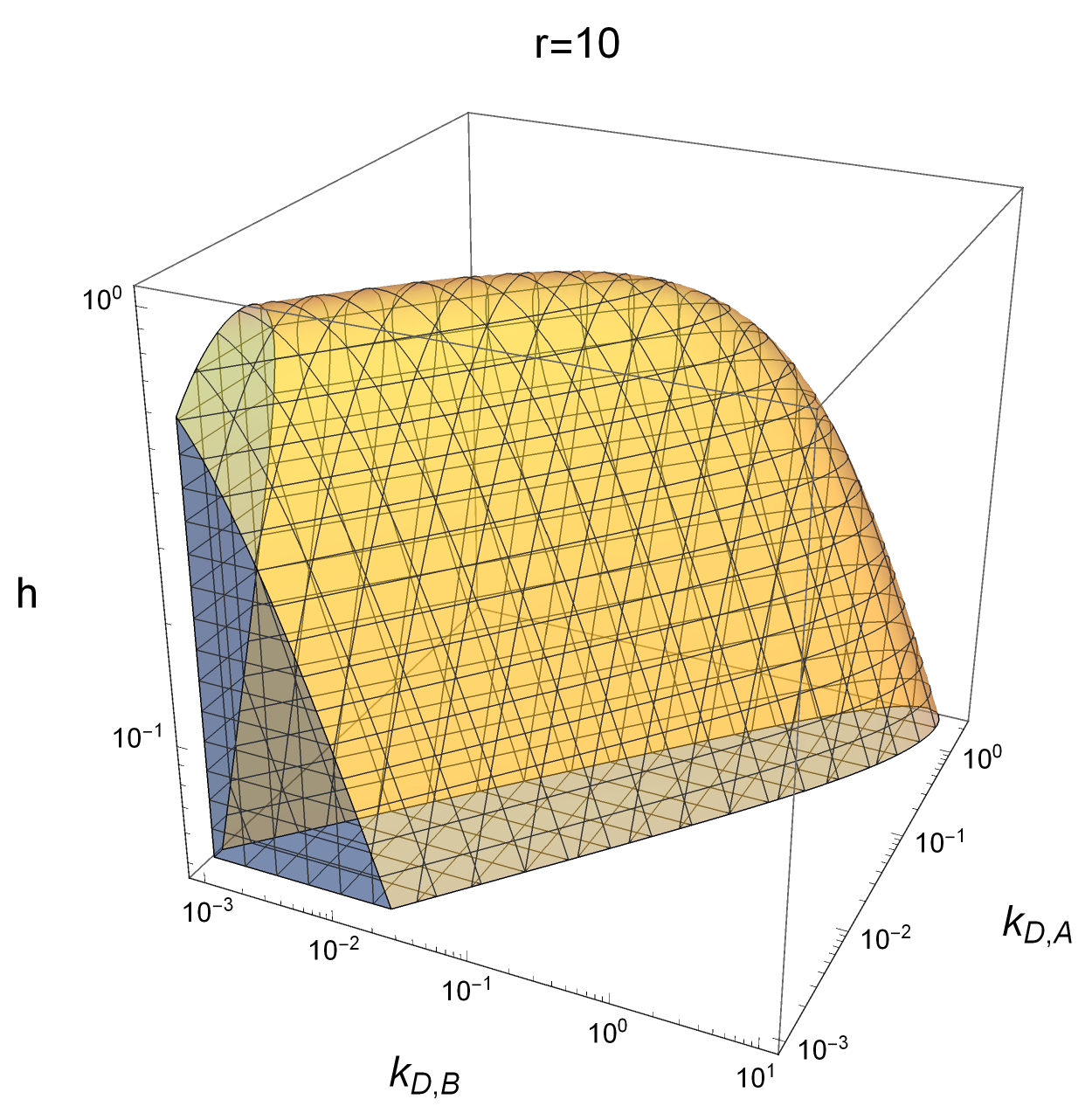}
\par\end{centering}
\caption{Parameter regions that satisfy the reaction requirements for diffusion
driven instability. The area inside the surfaces represent the parameter
space where diffusion driven instability can occur. \label{fig:reactionConstraints}}
\end{figure}

To have a diffusion driven instability the parameter values must satisfy
the following Routh-Hurwitz stability criterion
\begin{align}
f_{u}+g_{v} & <0\nonumber \\
f_{u}g_{v}-g_{u}f_{v} & >0.\label{eq:reactionConstraint}
\end{align}
For diffusion to destabilize this system, $df_{u}+g_{v}>0.$ These
inequalities imply that that either $f_{u}<0<g_{v}$ or $g_{v}<0<f_{u}$.
We found these criteria were satisfied when $k_{D,A}<2$, $k_{D,B}>10$,
and $r>1$ (Figure \ref{fig:reactionConstraints}).
 When performing the grid search with adaptive refinements, we set
the minimum value of $k_{D,A}$ and $k_{D,B}$ equal to 0.001, the
biological constraint obtained previously. We used integer values
for $r$ of 1, 2, 4, 6, 8, and 10. At $r=1$, there were no regions
that led to diffusion driven instability.

Although the constraints given by Equation \ref{eq:reactionConstraint}
are necessary for pattern formation, they are not sufficient. Whether
patterns can form and the number of possible modes also depends on
the size of the domain and the difference in diffusion rates between
the two classes of ILPs. For our purposes, the parameter $\gamma$
determines the size of the domain while $d$ gives the ratio of diffusion
rates. An estimate for $\gamma$ depends on the characteristic length
$L$ of the domain, the diffusion coefficient $D_{A},$ and the rate
of ILP degradation $k_{3}$ together determine the value of $\gamma$.
As discussed previously the protein degradation rate and diffusion
rates are likely between $10^{-3}$ and $10^{-6}$ $s^{-1}$ and 10
- 100 $\mu m^{2}$/s, respectively \citep{Belle2006,Cambridge2011,Milo2015}.
The characteristic length is equivalent to the length of the worm,
or 1.0 mm. This leads to possible values of $\gamma$ ranging from
0.01 to 100. Furthermore, we set a lower bound on the parameter $d$
based on the requirement that, for diffusion driven instability to
occur, $d$ must be greater than the critical diffusion value $d_{c}$
given by the roots of following equation (see \citep{Murray2000}
for derivation):
\begin{equation}
d_{c}^{2}f_{u}^{2}+2(2f_{v}g_{u}-f_{u}g_{v})d_{c}+g_{v}^{2}=0.
\end{equation}
For the example parameter set given (Table \ref{tab:parameters}),
the critical diffusion value $d_{c}$ is 2.97, implying the inhibitor
class $B$ must diffuse at least 3 times faster than the activator
class $A$. 

\begin{figure}
\centering{}

\subfloat[\label{fig:fig:dispersion-relationa}]{%
\noindent\begin{minipage}[t]{1\columnwidth}%
\includegraphics[scale=0.25]{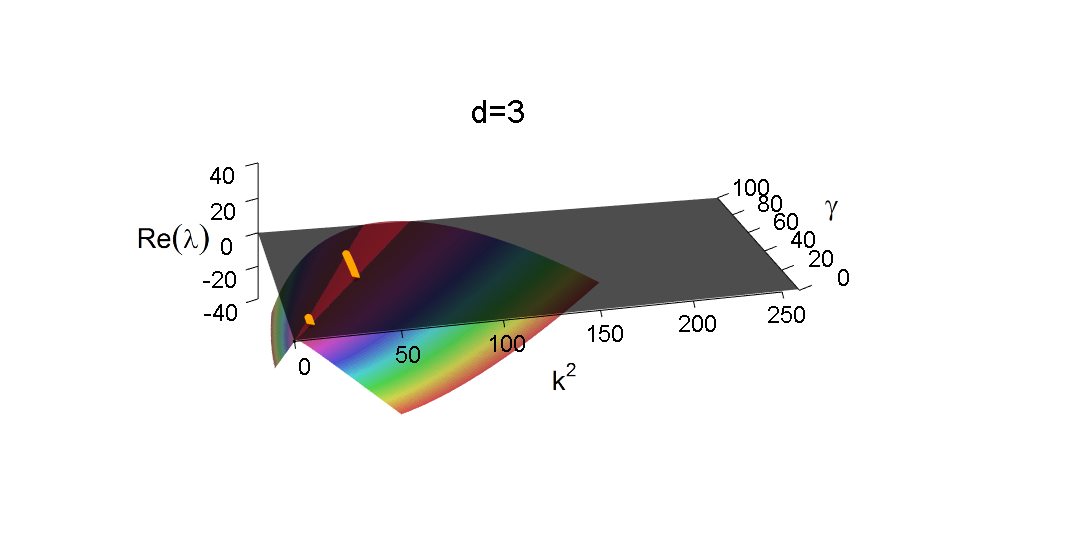}\includegraphics[scale=0.25]{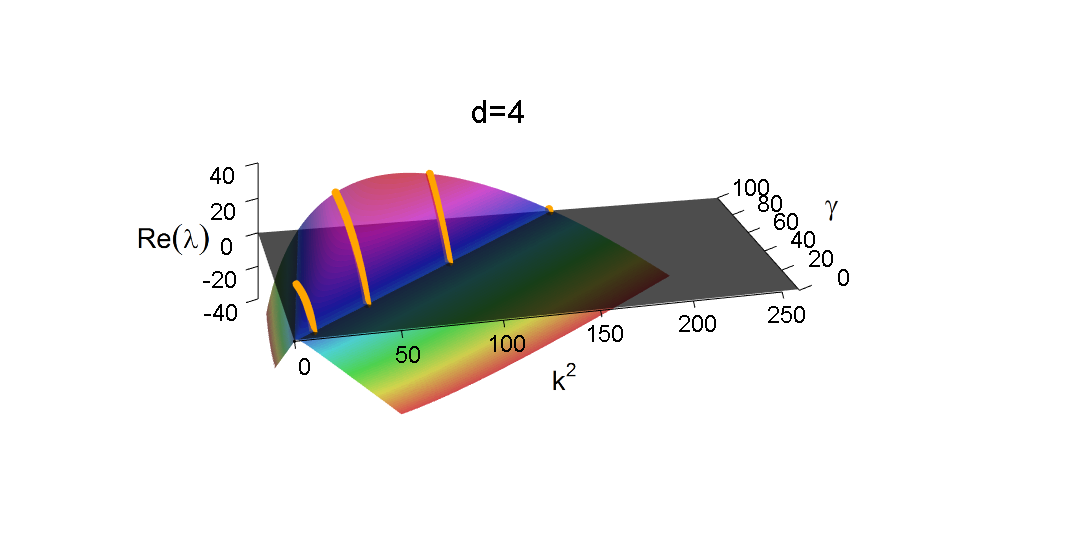}

\includegraphics[scale=0.25]{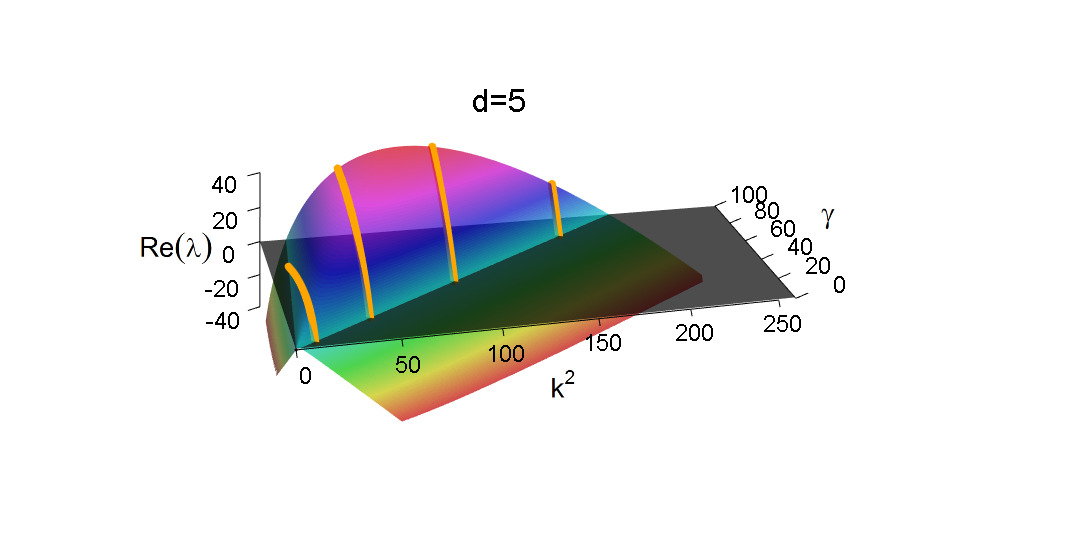}\includegraphics[scale=0.25]{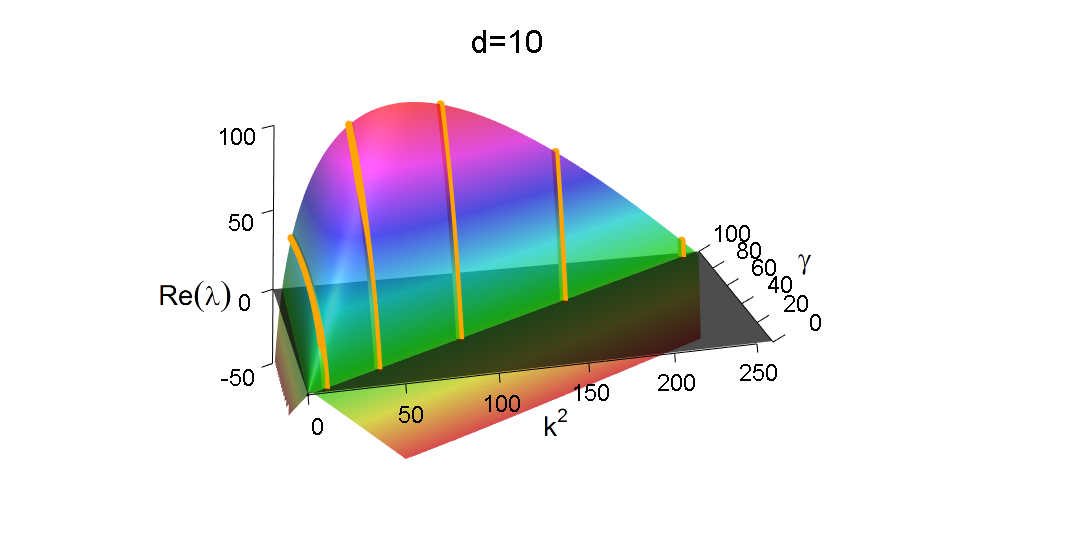}%
\end{minipage}}
\begin{centering}
\subfloat[\label{fig:fig:dispersion-relationb}]{\includegraphics[scale=0.4]{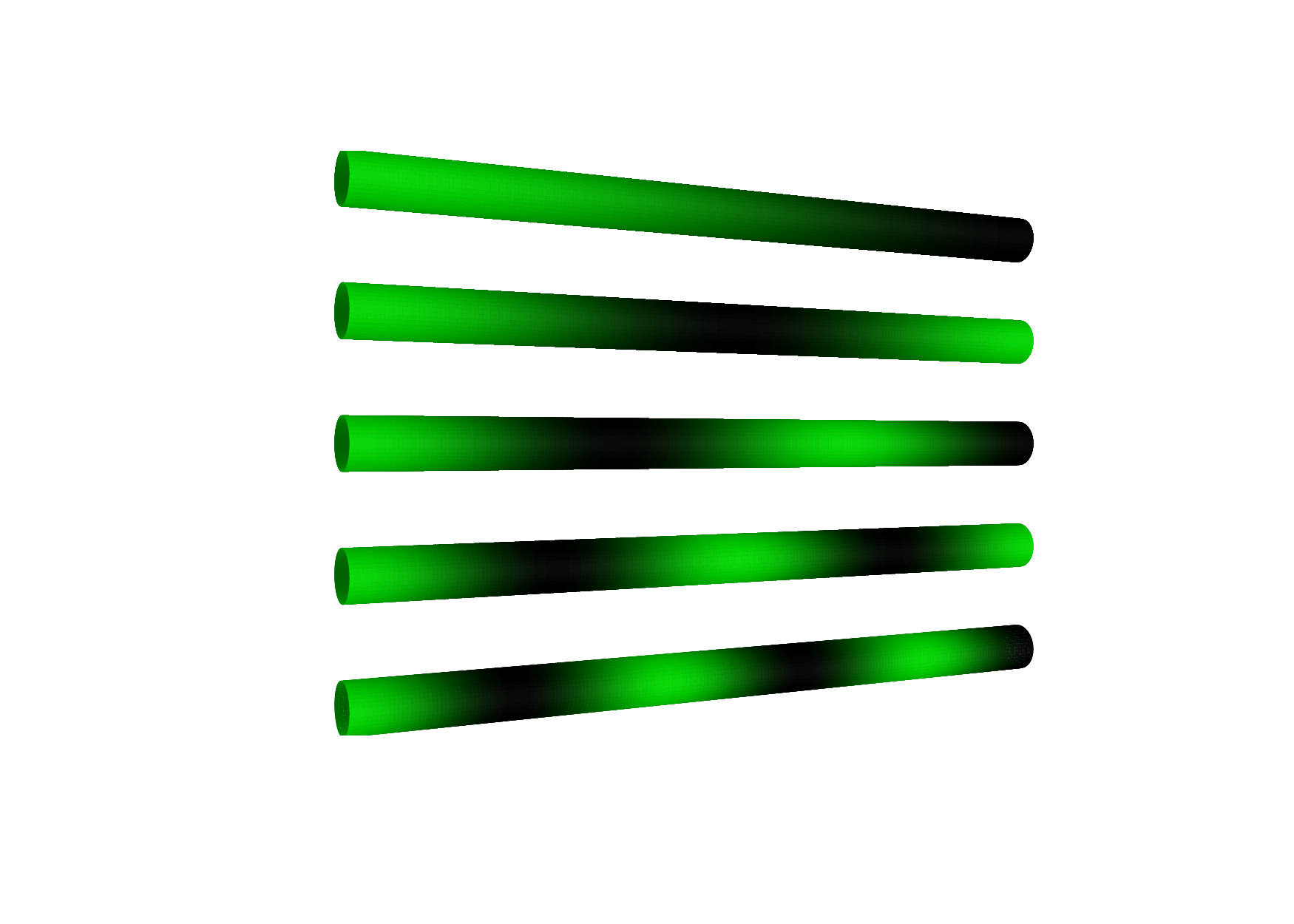} }
\par\end{centering}
\caption{The dispersion relation reveals that more modes are possible given
larger values of $\gamma$ and $d$. (a) The regions of the dispersion
relation curve (Equation \ref{eq:dispersionRelation}) with positive
values of $Re(\lambda)$ represent the possible wavenumber $k^{2}$
under which modes can grow. Due to finite domain constraints only
a discrete set of modes are possible, corresponding to discrete values
of $k^{2}$ (Equation \ref{eq:wavenumber}), shown with the yellow
lines in each figure. Each yellow line represents a different mode,
which from left to right correspond with $n=1,2,3,4,5$. (b) Visualization
of the five different possible modes with the lowest mode ($n=1)$
on top and the highest mode ($n=5$) on bottom. \label{fig:dispersion-relation}}
\end{figure}

We calculated the dispersion relation (Equation \ref{eq:dispersionRelation})
as a function of $\gamma$ at different values of $d$ to determine
how $\gamma$ and $d$ affect the number of possible modes (Figure
\ref{fig:dispersion-relation}). 
 For this calculation all other parameter values were obtained from
Table \ref{tab:parameters}. Larger values of $\gamma$ and $d$ lead
to a greater number of possible modes. In total when $d\le10$ and
$\gamma\le100$, one of five modes can arise (Figure \ref{fig:dispersion-relation}).
These modes correspond to setting $n=1,2,3,4,5$ and $m=0$ in Equation
\ref{eq:eigenfunctions}. For larger values of $\gamma$ the first
mode ($n=1$) eventually is no longer possible. In this analysis we
found that, when setting $d\le10$ and $\gamma\le100$, there were
no modes possible in which $m>0$. This implies that fluctuations
in the HSP expression only occur along the length of the worm and
not along the circumference of the worm.

\begin{figure}
\includegraphics[scale=0.6]{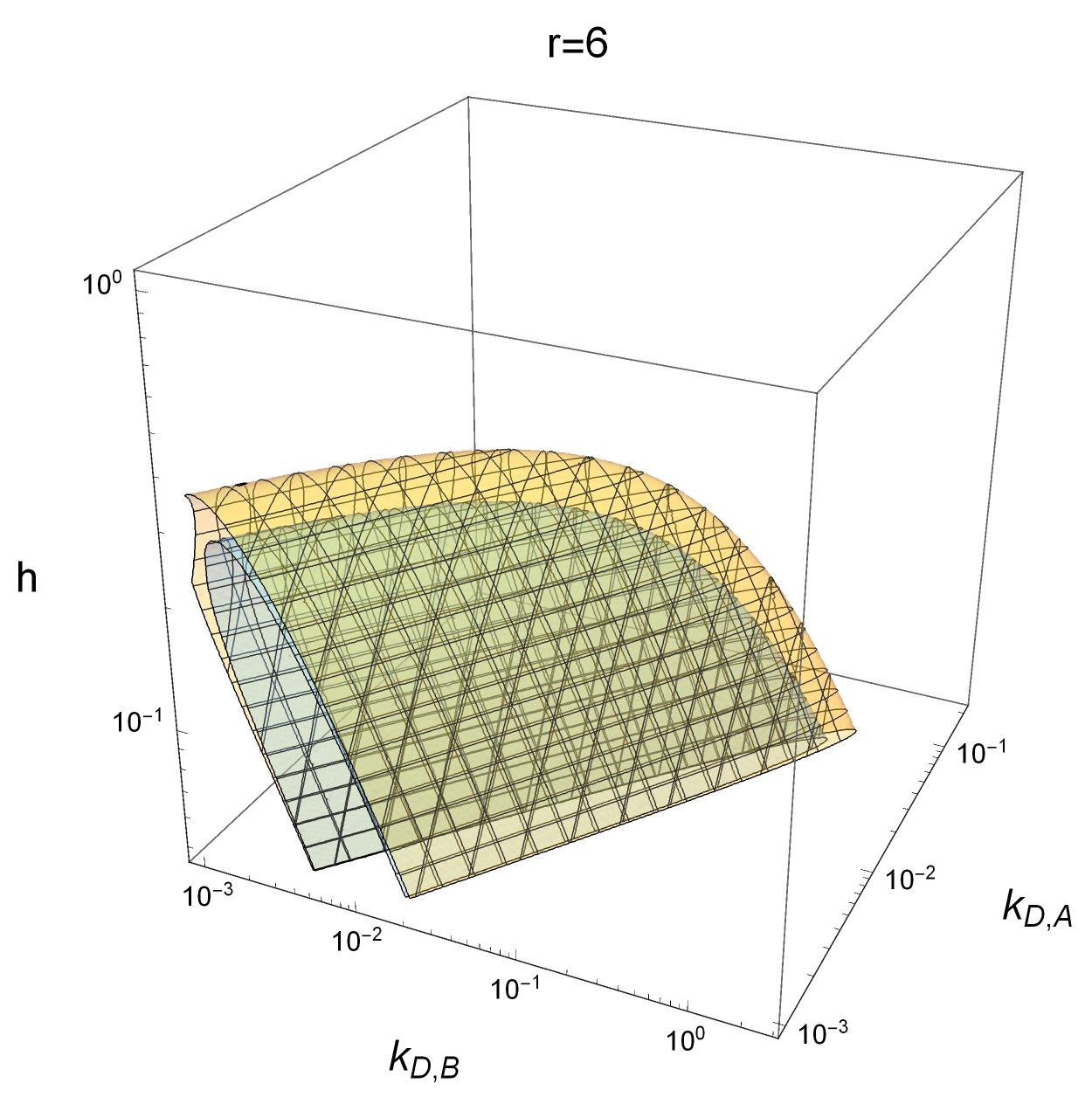}\includegraphics[scale=0.6]{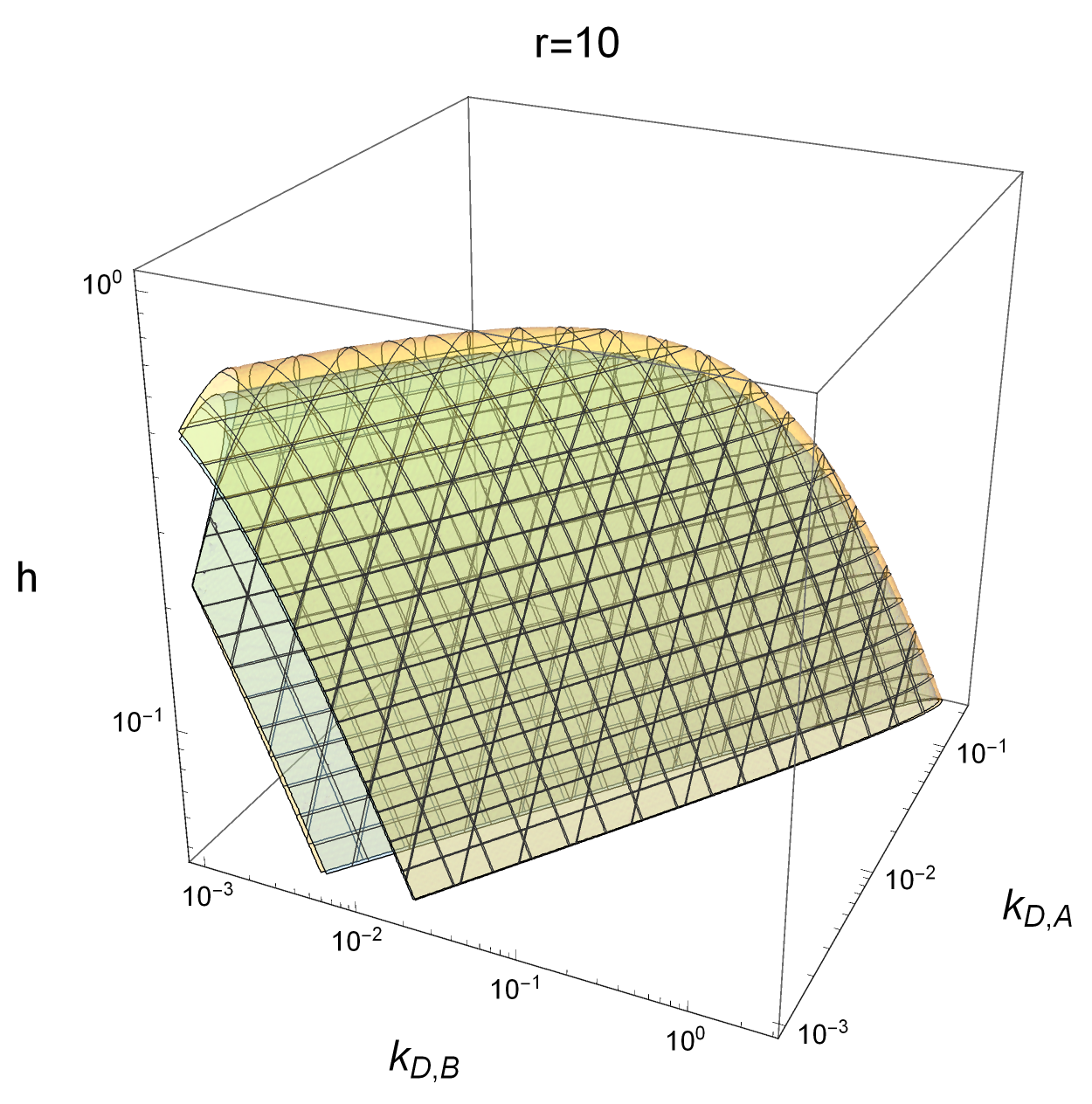}\caption{Total number of possible modes as a function of parameter values.
In the larger orange shaded region only one mode is possible and in
the smaller blue shaded region at least two modes are possible. \label{fig:modeNumber}}
\end{figure}

To determine the dimensionless dissociation constants with more precision,
we examined which parameter values resulted in patterned modes and
how many modes were possible. We did this by using the parameters
given in Table \ref{tab:parameters} but varied $r$, $k_{D,A}$,
$k_{D,B}$, and $h$ to determine the number of possible stable modes
as a function of the parameter values (Figure \ref{fig:modeNumber}).
 Given the constraints on the parameter values specified previously,
to obtain at least one pattern formation mode $k_{D,A}$ must be less
than 0.2, and $k_{D,B}$ must be less than 2.5 and $h$ must be less
than 1. The parameter values become further constricted when we require
that the system have at least two modes that grow with time. 

\subsection{Worms fall into two distinct groups based on HSP expression\label{subsec:dataResults}}

\begin{figure}
\subfloat[\label{fig:projectionAnalysisa}]{\begin{centering}
\includegraphics[scale=0.8]{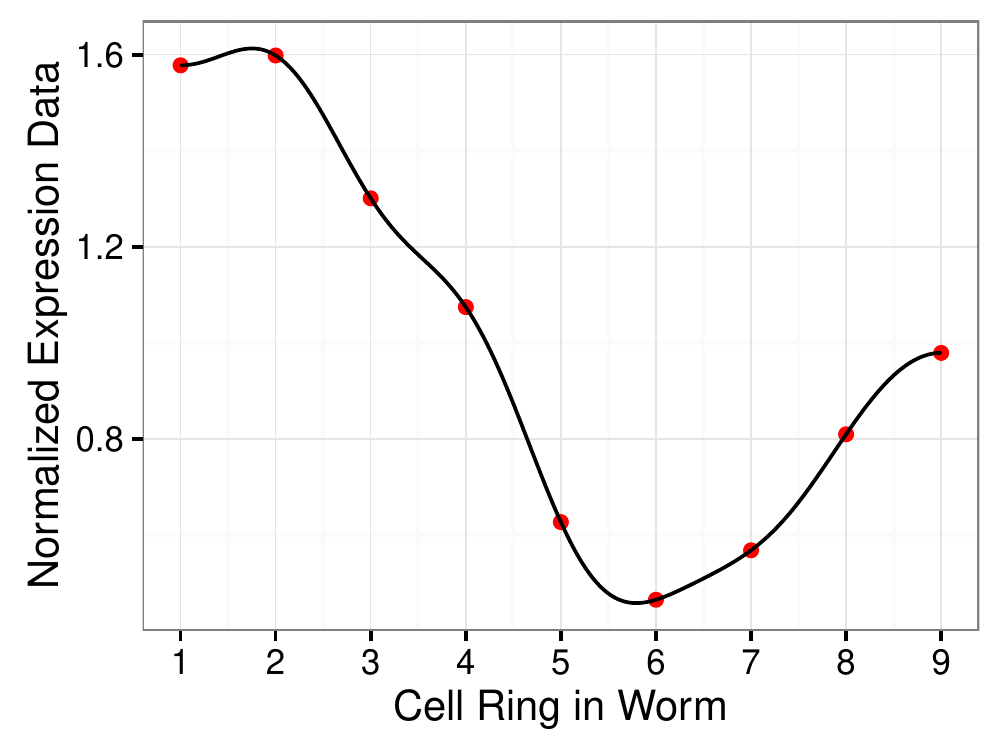}
\par\end{centering}
}\subfloat[\label{fig:projectionAnalysisb}]{\centering{}\includegraphics[scale=0.8]{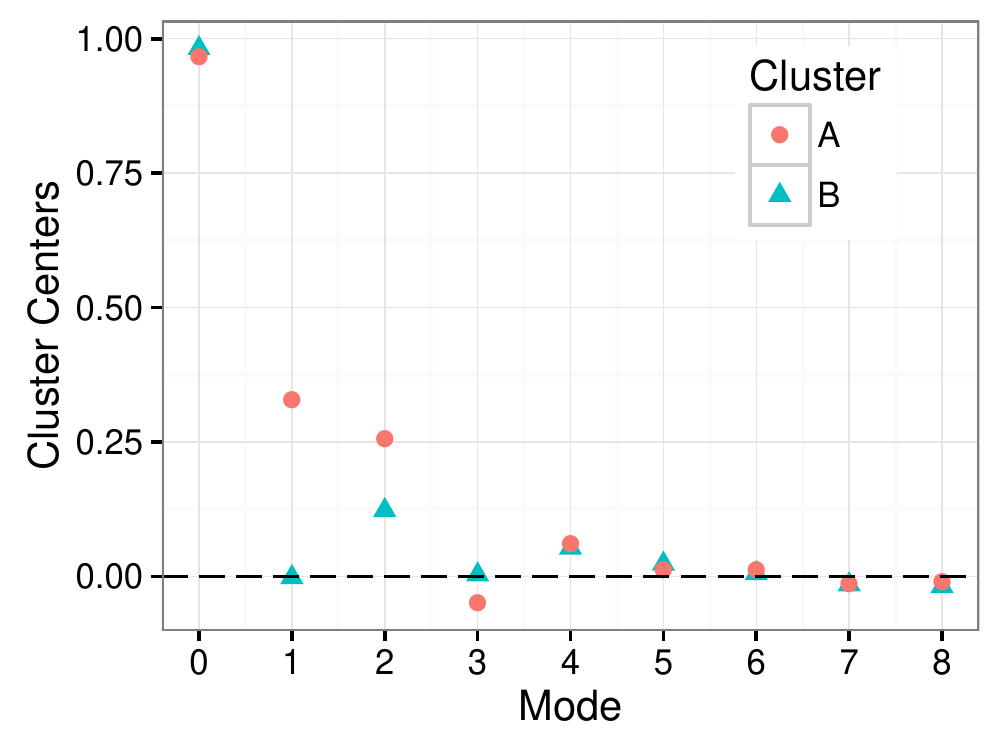}\label{fig:ExpansionCoefficients}}

\subfloat[\label{fig:projectionAnalysisc}]{\begin{centering}
\includegraphics[scale=0.8]{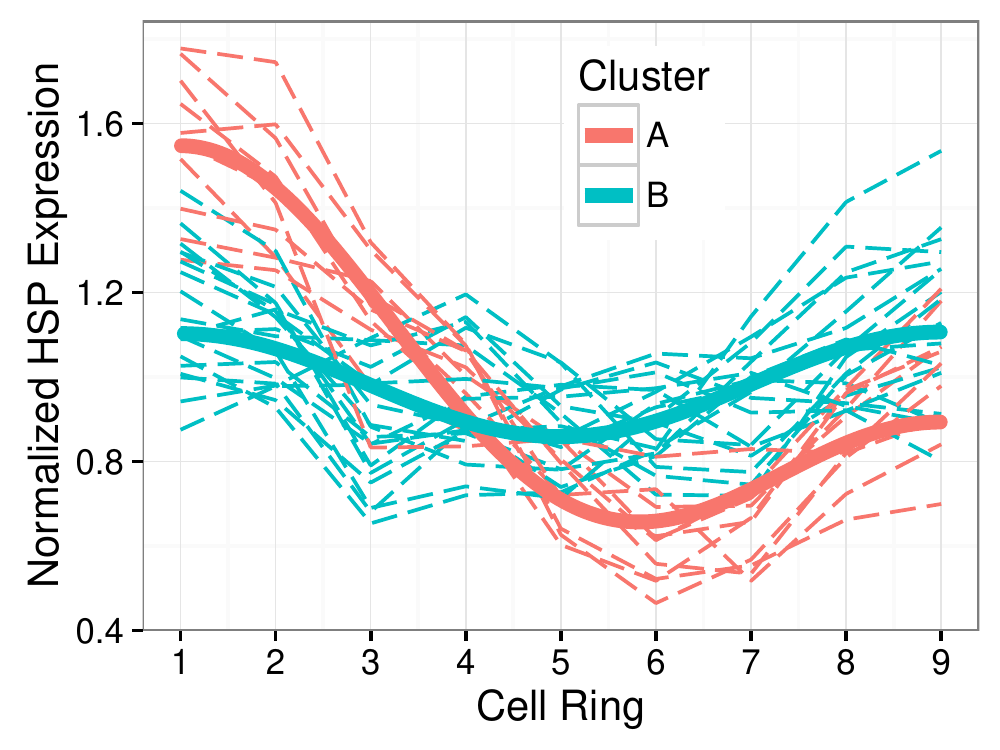}
\par\end{centering}
}\subfloat[\label{fig:fig:projectionAnalysisd}]{\begin{centering}
\includegraphics[scale=0.8]{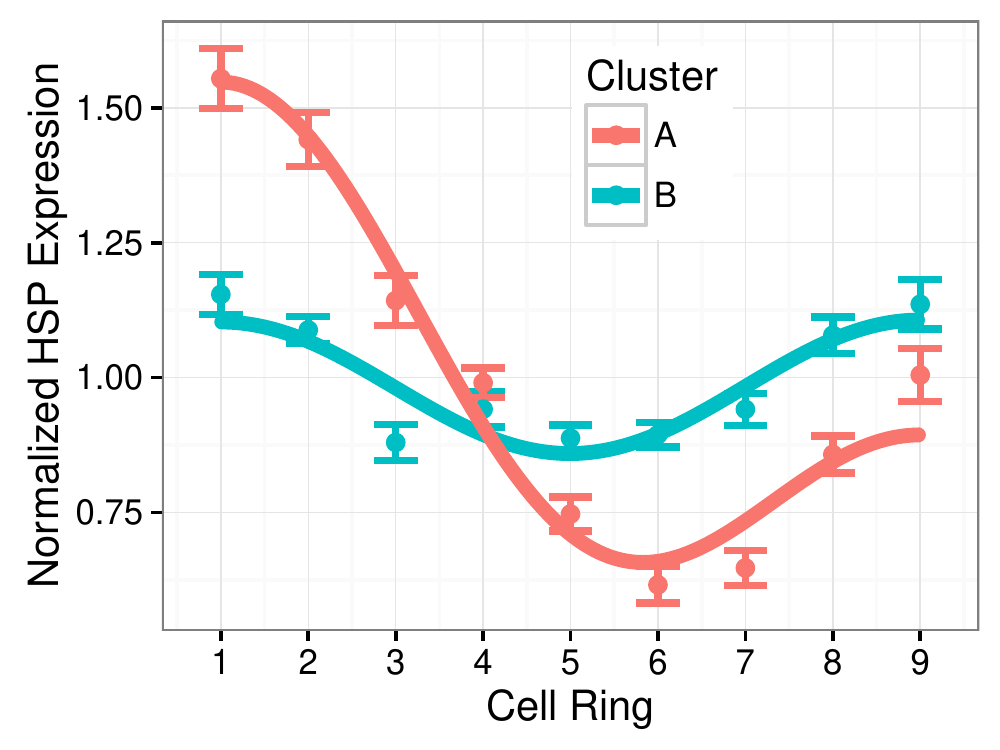}
\par\end{centering}
}

\caption{Clustering analysis reveals two distinct groups of worms. (a) Example
individual worm data with interpolation using cosine modes weighted
by expansion coefficients. (b) Cluster centers following a k{[}means
analysis where clustering was done using the expansion coefficients
for each cosine mode. (c) Individual worm data (thin dashed lines)
compared with predicted profiles (thick solid lines) using summation
of the dominate modes ($n$ = 0, 1 and 2 for Cluster A; $n$ = 0 and
2 for Cluster B). (d) Same as (c) but now showing the mean and standard
error of the individual worm data. \label{fig:projectionAnalysis}}
\end{figure}

To determine if the modes resulting from the hypothesized diffusion
driven instability can explain experimental data on HSP expression
patterns in \emph{C. elegans}, we performed a clustering analysis.
Worms were grouped using the expansion coefficients obtained from
projecting trigonometric modes onto the data (Figure \ref{fig:projectionAnalysisa}).
 This analysis reveals that \emph{C. elegans} fall into two distinct
groups based on HSP expression patterns. Using bootstrap methods,
we found that grouping the worms into two clusters had strong support
with mean Jaccard similarity scores of 0.92 and 0.89. However, when
the number of groups was increased to three, this support decreased
to 0.51, 0.79, and 0.60 for the three clusters. 

Using the two cluster approach, we determined the dominate modes for
each cluster (Figure \ref{fig:ExpansionCoefficients}). Clearly the
$n=0$ mode contributes significantly since this represents the averaged
signal. For Cluster A the first and second mode also appear to have
large contributions, and for Cluster B the second mode appears to
have a large contribution. To determine if these specified modes are
sufficient to explain the data, we overlayed the predicted spatial
profiles using only these modes onto the the individual worm data
(Figure \ref{fig:projectionAnalysis}c) and the mean worm data (Figure
\ref{fig:projectionAnalysis}d). The predicted profiles using only
the the specified dominate modes demonstrate a good match with the
data.

\subsection{Pattern formation in the Simulations}

Our previous analysis reveals that a linearized version of the mathematical
model produces stable oscillatory solutions. This result, however,
does not guarantee that patterns will emerge from the full nonlinear
model. Accordingly, we generated computational simulations of the
nonlinear system. This allowed us to both check the validity of our
analysis and verify that the model can reproduce the data on \emph{hsp-16.2}
transcription in \emph{C. elegans.} We performed a series of computational
simulations of the full PDE system where we varied the values of $\gamma$
and $d$ and kept other parameter values constant (see Table \ref{tab:parameters}
for values used). These simulations provide evidence supporting the
existence of stable patterns, where the steady state is represented
by a single dominate mode. We next compared these results to the clustering
analysis presented previously. The model provides an explanation for
how the pattern of a single dominate mode observed in Cluster B emerges
(see Figure \ref{fig:projectionAnalysis}). However, the steady-state
results do not explain the pattern of two equally dominate modes observed
in Cluster A. We hypothesize that Cluster A is in a quasi-steady state
where two modes are competing for dominance. 

\begin{figure}
\centering{}\subfloat[]{\centering{}\includegraphics[scale=0.45]{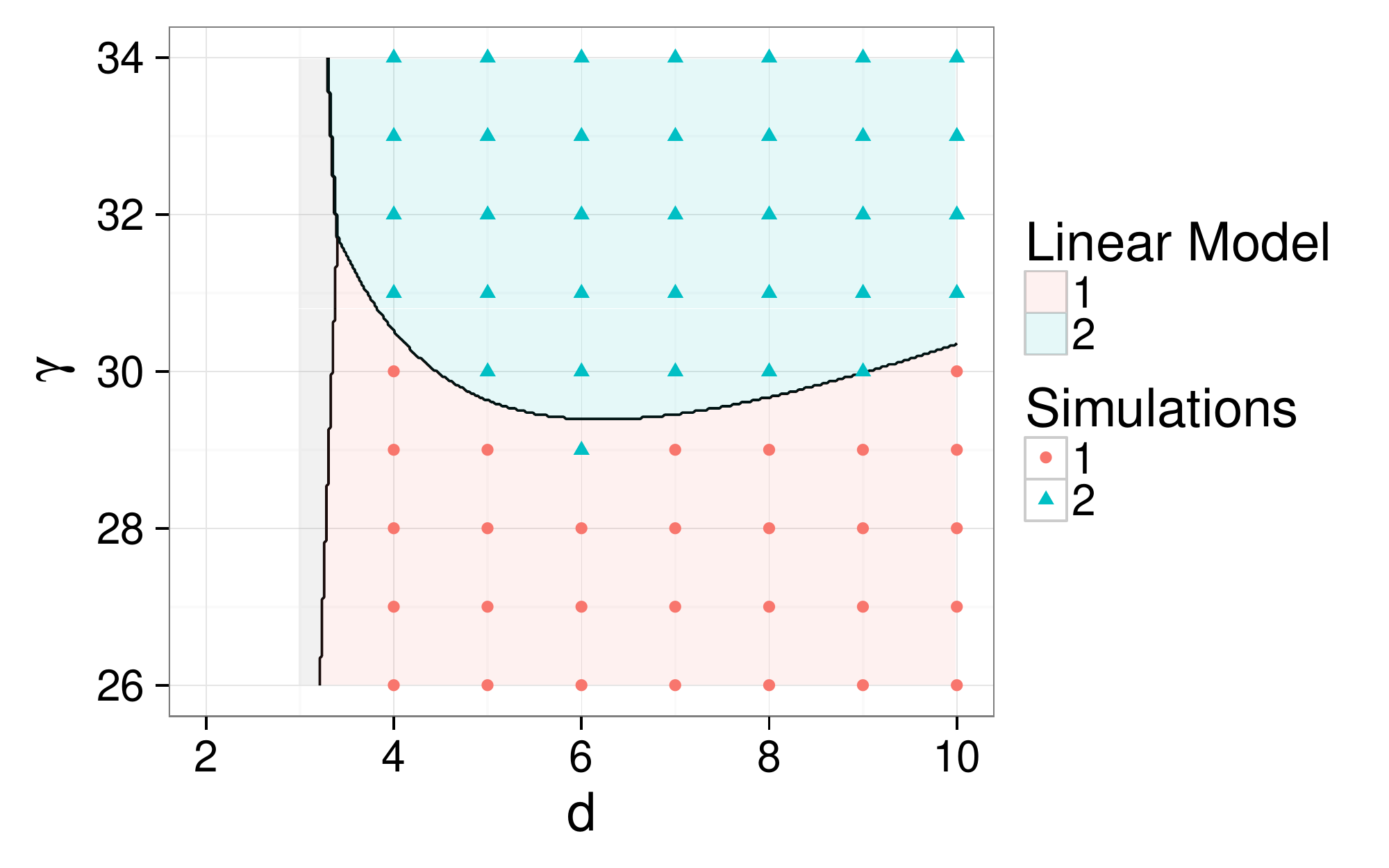}}\subfloat[]{\centering{}\includegraphics[scale=0.45]{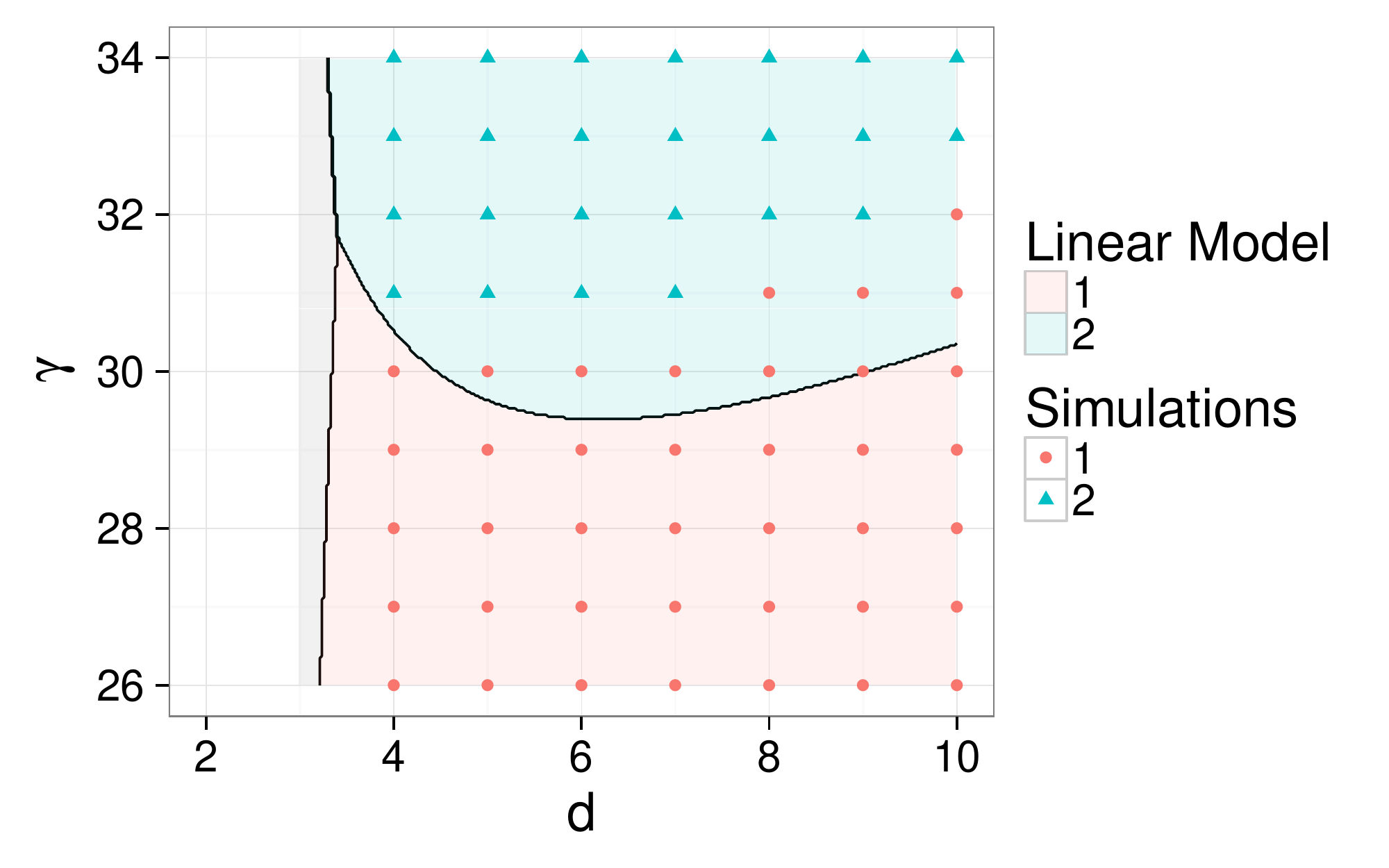}}\caption{ Comparison of the steady states obtained in simulations with predictions
from the linearized model. (a) and (b) represent sets of simulations
ran using two different initializations of Gaussian noise around the
steady state. Each dot represents a single simulation of the full
non-linear model and the color of the dot represents the dominate
mode that the system eventually enters. The shaded regions represent
the fastest growing mode as predicted by the linearized model. The
linear model prediction roughly matches the dominate mode obtained
using simulations of the full non linear model. However, the exact
transition point from a first mode to second mode dominance depends
on the initial conditions.\label{fig:modePrediction}}
\end{figure}

To determine if a quasi-steady state is possible, we examined whether
we can predict the transition from a first mode dominate to a second
mode dominate state based on the parameter values. In turn, we predict
a quasi-steady state representing a combination of the first two modes
may exist near this transition region. While our analysis of the linearized
model predicts that several modes could exist, it does not predict
which mode will dominate. There are analytical tools to perform such
predictions and, for low-dimensional modes, they typically yield that
the dominant mode will be the fastest growing mode or the mode for
which $Re(\lambda)$ is maximized \citep{Murray2000}. To see if this
method could be used to predict the transition point from a first
mode to a second mode dominate state, we compared the fastest growing
mode to simulation results for the steady state (Figure \ref{fig:modePrediction}).
 When the model is initialized using random noise around the steady-state
for the reactions, we are able to roughly predict the transition region
as a function of $\gamma$ and $d$. However, the exact transition
point is sensitive to the initial conditions of the simulation. 

\begin{figure}
\begin{centering}
\includegraphics[scale=0.25]{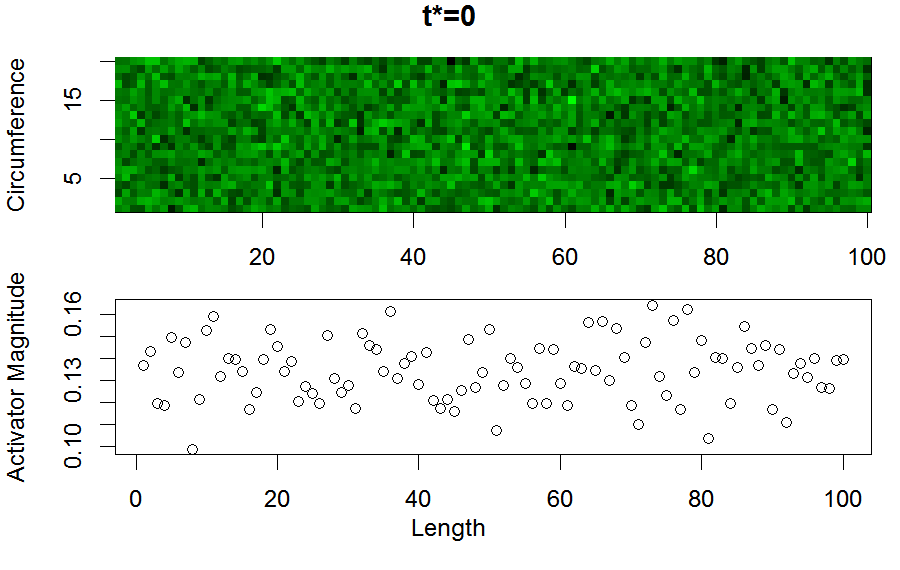}\includegraphics[scale=0.25]{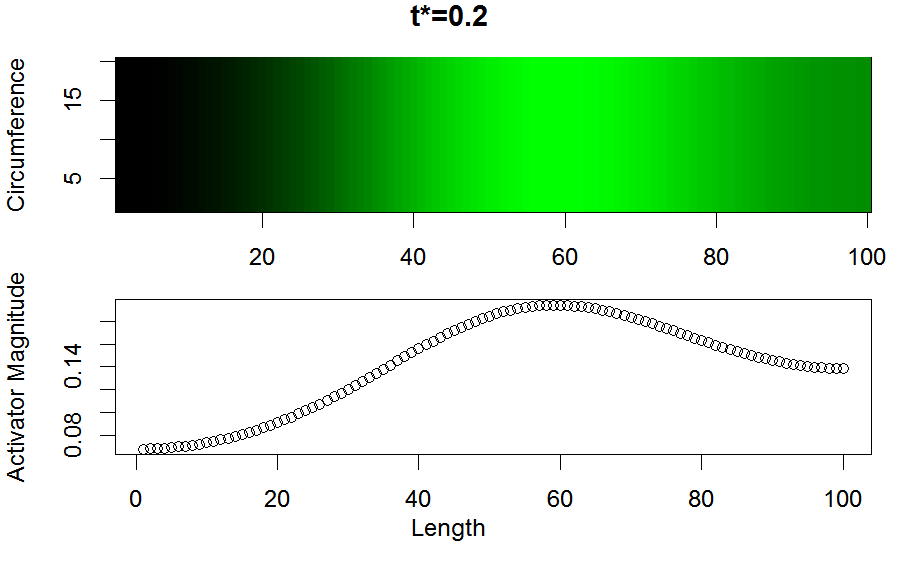}
\par\end{centering}
\begin{centering}
\includegraphics[scale=0.25]{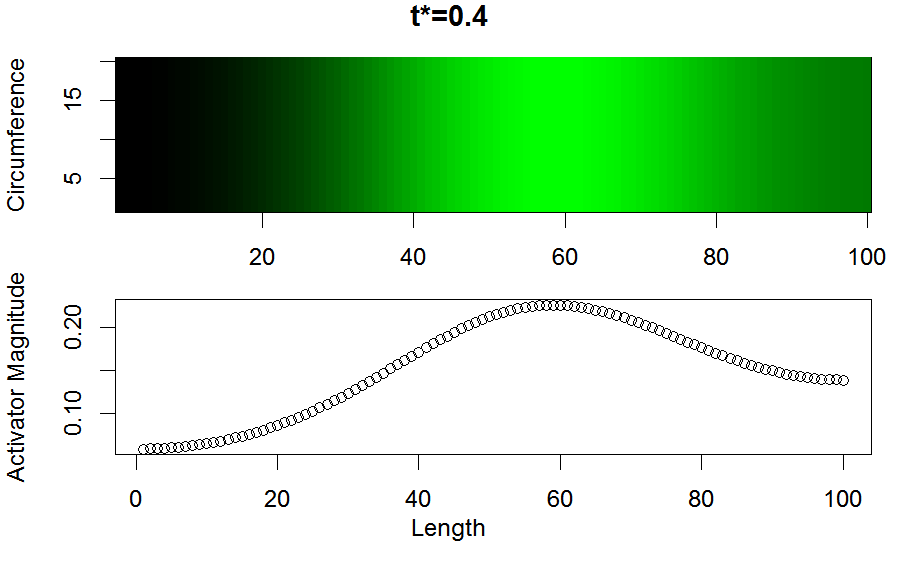}\includegraphics[scale=0.25]{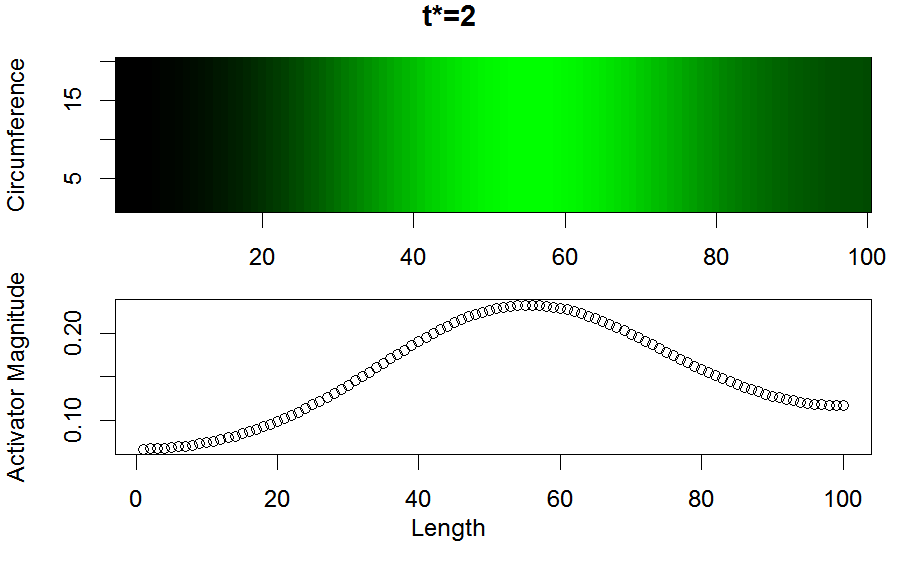}
\par\end{centering}
\caption{Simulation initialized with Gaussian noise that enters a quasi-steady
state. Showing simulation results at four different values of $t^{*}$.
The activator magnitude shown represents the top horizontal line shown
in the domain. \label{fig:simulation_a}}
\end{figure}

\begin{figure}
\begin{centering}
\includegraphics[scale=0.25]{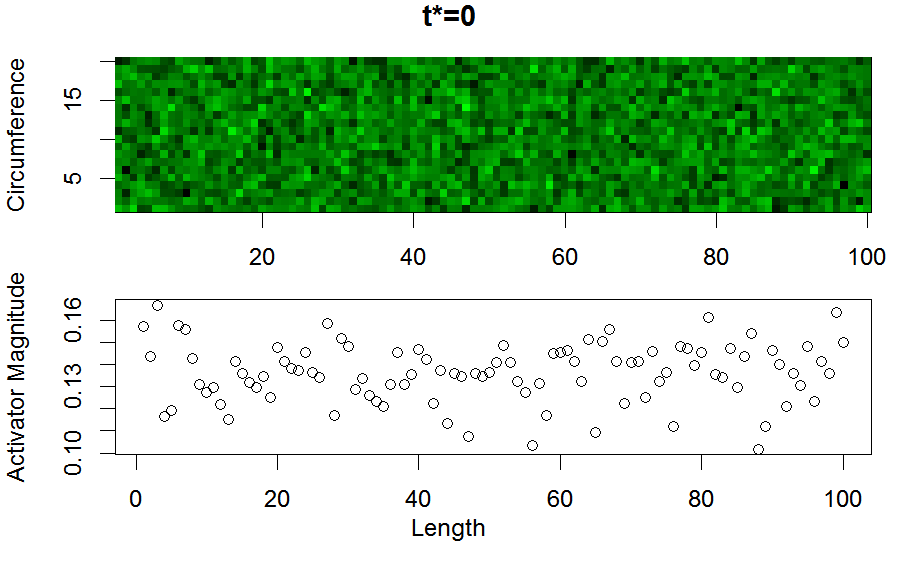}\includegraphics[scale=0.25]{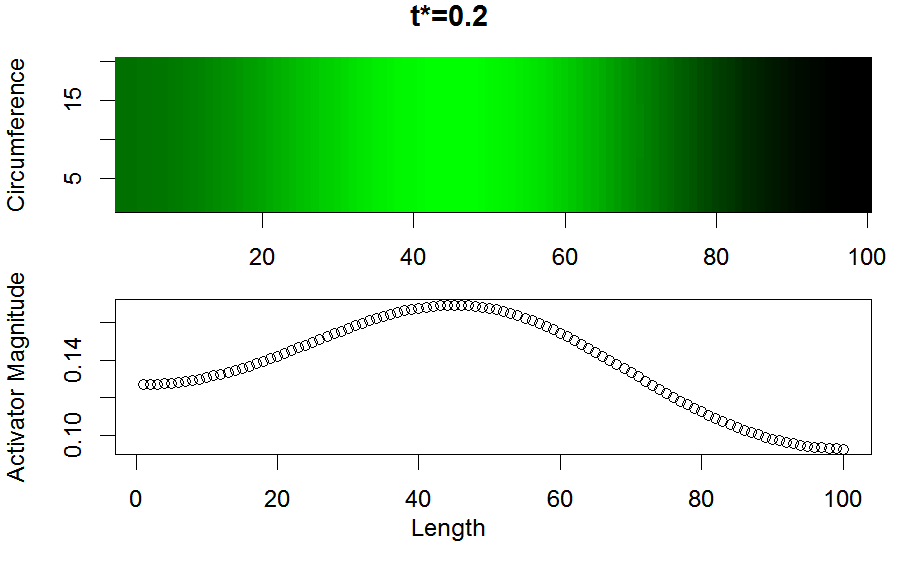}
\par\end{centering}
\centering{}\includegraphics[scale=0.25]{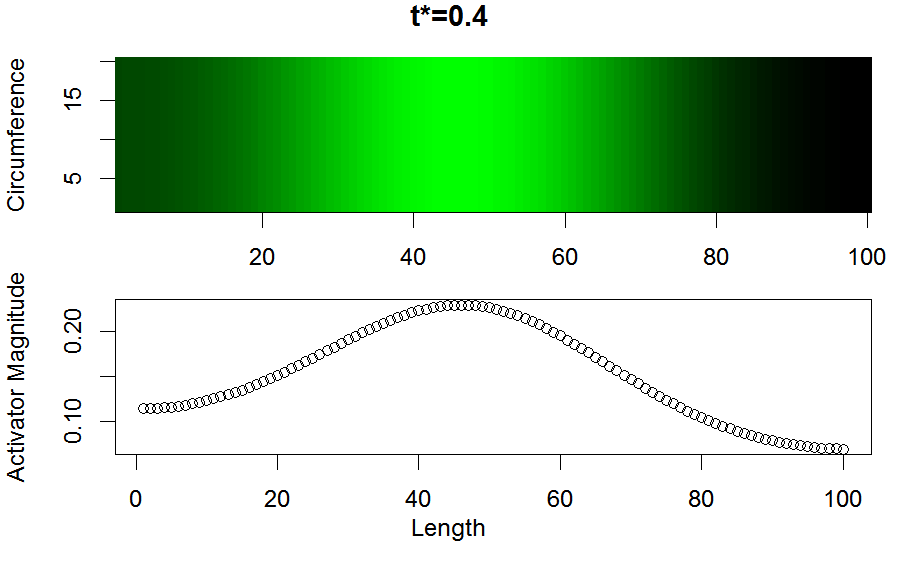}\includegraphics[scale=0.25]{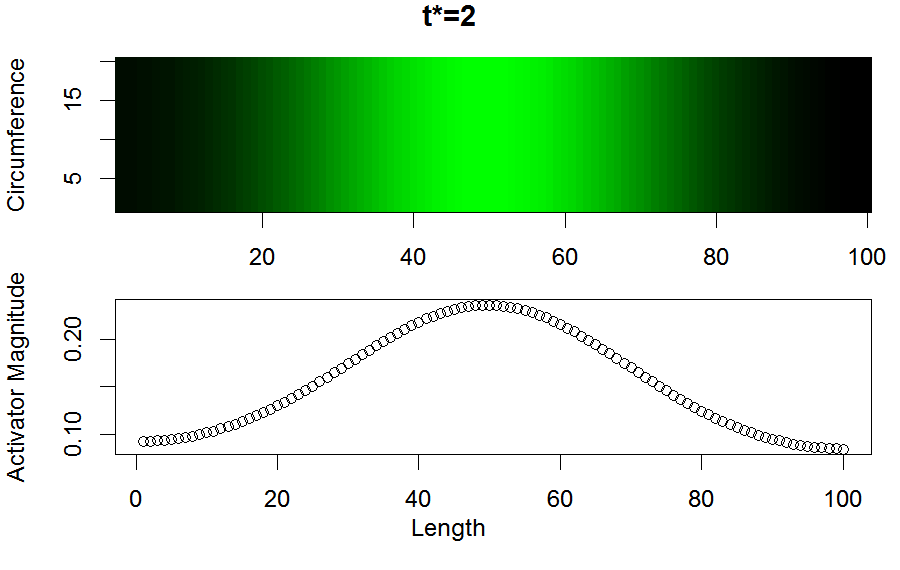}\caption{Simulation initialized with Gaussian noise that has entered a steady-state
mode by $t^{*}=2$. Showing simulation results at four different values
of $t^{*}.$ The activator magnitude shown represents the top horizontal
line shown in the domain. \label{fig:simulation_b}}
\end{figure}

To further explore this result we performed simulations at points
in parameter space near the transition region. Specifically, we set
$\gamma=30$.2 and $d=5$. At this point, simulations initialized
using Gaussian noise around the steady-state showed that the second
mode is dominate. We found with this parameter combination that the
system sometimes enters a quasi-steady state (Figure \ref{fig:simulation_a}
and \ref{fig:simulation_b}).
 However, the progression into this state is sensitive to the specific
Gaussian noise present in the initialization. For example, in one
initialization the system appears to still be in a quasi-steady state
at $t^{*}=2.0$ (Figure \ref{fig:simulation_a}), while in a second
initialization the system has entered the steady state at $t^{*}=2.0$
(Figure \ref{fig:simulation_b}). 

\begin{figure}
\centering{}\includegraphics[scale=0.25]{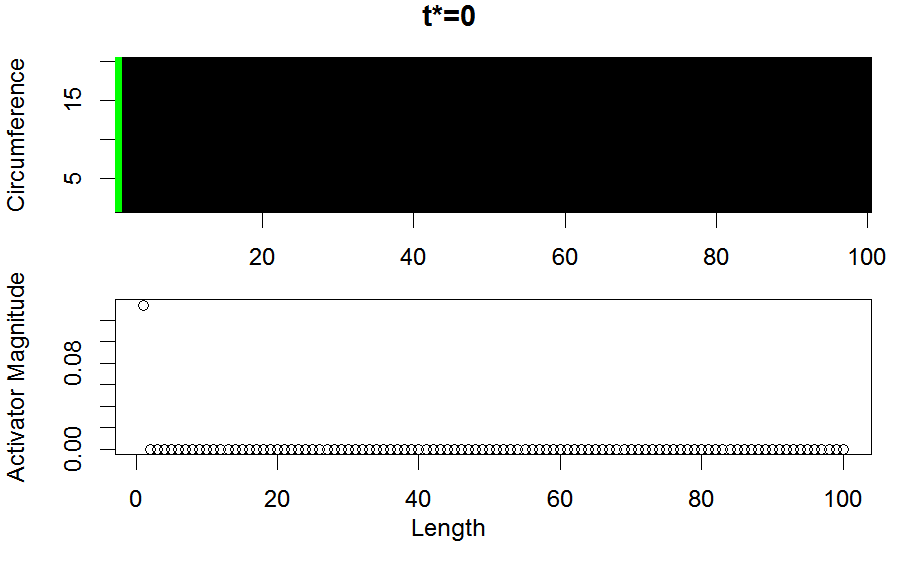}\includegraphics[scale=0.25]{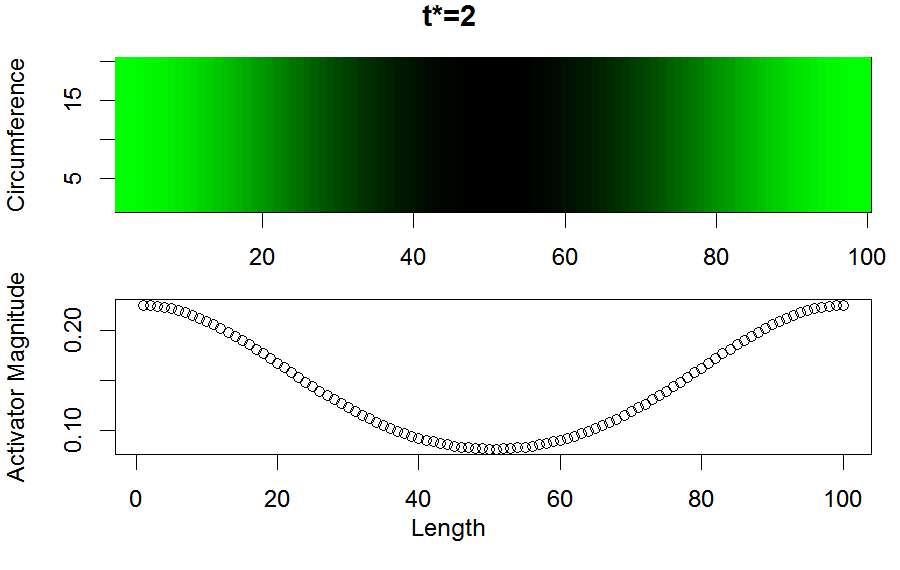}\caption{Results of simulation initialized using left constant conditions.
(Left) Condition used to initialize simulation. (Right) Simulation
results where $\gamma=37$ and $d=4$. All other parameters are given
in Table \ref{tab:parameters}. The activator magnitude shown represents
the top horizontal line shown in the domain. \label{fig:simulation_c}}
\end{figure}

We next performed simulations where the pattern was initialized at
the left end of the domain. Biologically this corresponds to a release
of ILPs at the anterior region of the worm's intestine. We found that
at multiple values of $\gamma$ and $d$ the observed pattern orientation
was accurately reproduced (an example is shown in Figure \ref{fig:simulation_c}).
 The simulation predicts increased \emph{hsp-16.2} transcription at
the head and the tail of the worm and this matches what is observed
experimentally. This is in contrast to the simulations initialized
using Gaussian noise across the domain (Figure \ref{fig:simulation_a}
and \ref{fig:simulation_b}). These simulations predict reflected
versions of the pattern may arise (i.e., decreased transcription of
\emph{hsp-16.2} at either end of the domain compared with the middle).
Thus, the initial conditions may explain why only one orientation
of mode 1 and 2 are observed experimentally.

\section{Discussion}

We developed a reaction diffusion model that predicts experimentally
observed patterns in the transcription of \emph{hsp-16.2}, a biomarker
of aging \citep{Mendenhall2015,Rea2005}. This work provides insight
into the mechanism of the non-cell autonomous heat shock response
in the nematode \emph{C. elegans}. We hypothesized that the transcription
of \emph{hsp-16.2} is dependent on a complex interplay of insulin
like peptides signaling at a multicellular level. This complex system
warrants the use of mathematical tools to understand how changes in
the system affect the total levels of \emph{hsp-16.2} transcription.
Furthermore, since \emph{hsp-16.2} transcription is correlated with
lifespan, the development of a mathematical model that connects thermal
injury to intestinal \emph{hsp-16.2} transcription allows us to study
how alterations to kinetic or diffusive parameters in this system
might be related to aging.

We derived constraints on the values of four parameters in the system
by combining biological intuition with the requirement for diffusion
driven instability. To obtain patterns, the reaction parameters must
satisfy the Routh-Horwitz conditions. Furthermore, parameter values
are constrained by the spatial and diffusive properties of the system,
including the difference in the diffusion rate between the two classes
of ILPs and the domain size. Ultimately, when setting the minimum
value of $h$ to 0.05, and the maximum value of $r$ to 10, to obtain
patterned modes $k_{D,A}$ must be between 0.001 and 0.2 while $k_{D,B}$
must be between 0.001 and 2.5 (see Table \ref{tab:parameters} for
the values of other parameters). These dimensionless dissociation
constants are equal to the actual dissociation constant multiplied
by the ILP decay/production ratio ($k_{3}/k_{1}$, $k_{3}/k_{4}$).
Although the actual dissociation constants for the DAF-2 agonists
and antagonists are currently unknown, they are likely on the order
of 10-100 nM \citep{Subramanian2013}, leading to an ILP decay/production
ratio range of $10^{-5}$ to $10^{-1}$ $nM^{-1}.$We also obtained
parameter constraints related to the feedback strength in the system.
The parameter $h$ represents the strength of the feedback; as $h$
increases more stimulus is required for the same response size. Thus,
an upper bound on the value of $h$ of 1.0 implies that a minimal
feedback strength is required for pattern formation to occur. We found
no lower bound on the value of $h$, suggesting that there is not
a maximal feedback strength. The Hill coefficient $r$, which determines
the sharpness of the feedback response must be greater than one, suggesting
some cooperativity in the system may exist. There are several signaling
motifs that could lead to a Hill coefficient greater than one \citep{Zhang2013a}.
One such process involves non-processive multisite protein phosphorylation
by a single kinase. The nuclear translocation of DAF-16 is potentially
blocked when DAF-16 is phosphorylated by AKT at one or multiple sites
\citep{Lin2001}. Thus, it may be the requirement for multi-site phosphorylation
of DAF-16 that leads to the large Hill coefficient predicted by the
model.

To derive these parameter constraints, the values of three other parameters
were set to constant values. By setting $a$ and $b$ equal to one,
we assumed ILP transcription was completely dependent on DAF-16 nuclear
localization. However, this is likely not the case since there could
be other transcription factors that affect ILP transcription. If $a$
and $b$ were less than one, this would imply that there is a baseline
level of ILP production without DAF-16 nuclear localization. Values
of $a$ and $b$ greater than one are not biologically possible, since
this would lead to negative production rates. The third parameter
value $c$ was set equal to one, implying that the two classes of
ILPs decay at the same rate. This is an assumption that warrants future
investigation. Very little is currently known about ILP degradation
in \emph{C. elegans. }In higher level organisms, such as rats and
humans, insulin is typically endocytosed by cells with insulin-receptors
and is degraded within the cells \citep{Bai1995,Duckworth1998}. These
processes likely depend on specific properties of the insulin-like
peptides such as specific and non-specific binding affinities that
determine the likelihood of endocytosis and subsequent degradation.

The model presented here provides a mechanistic explanation for the
data obtained on \emph{hsp-16.2} expression patterns in \emph{C. elegans.
}A previous study found that \emph{hsp-16.2} transcription in intestinal
cells was greatest in the anterior region, followed by the posterior
region, and finally least in the middle of the worm \citep{Mendenhall2015}\emph{.}
We sought to use the developed model to provide an explanation for
this spatial profile. Since the mean spatial profile of the 28 worms
studied did not clearly match one of the modes predicted by the model,
we hypothesized that the worms fell into multiple groups with different
expression patterns. Using a clustering analysis, we found that the
worms could be divided into two groups. One of these groups had a
clear single dominate mode; however, the second group had two modes
that contributed equally to the spatial profile. Due to a lack of
bootstrap support, we did not attempt to divide the worms up into
more than two groups. 

The cluster with two dominate modes may represent worms that have
not yet reached a steady state in their \emph{hsp-16.2} transcription
spatial profiles. To explore this possibility we performed simulations
to determine if the system could get temporarily stuck in a state
between the first two modes. We found that a reasonable method for
predicting whether the first or second mode would dominate is to determine
what mode grows the fastest according to the linearized model. We
predicted that if two modes grow at a roughly equal pace, then the
system may reside in a quasi-steady state that represents a superpositioning
of the two modes. In our simulations we found that this was indeed
the case (Figure \ref{fig:simulation_a} and \ref{fig:simulation_b});
however, the dynamics of the system ultimately depended on the initial
conditions. This dependency could explain why some worms appear to
have reached a steady state 24 hours after heat shock and other worms
have not.

The mathematical model presented demonstrates how patterns can emerge
but represents a simplification of the actual signaling pathways within\emph{
C. elegans}. There are over 40 ILPs that have been identified in \emph{C.
elegans, }and the action of each ILP does not necessary fit precisely
into one of the two classes of ILPs included in the model. For instance,
in the model DAF-16 only has a stimulatory effect on the production
of ILPs. In reality, DAF-16 increases the production of some ILPs
and decreases the production of others. This leads to ILP agonists
that are both self-activating (e.g., INS-7) and self-inhibiting (e.g.,
INS-6) \citep{Hua2003,Murphy2003a}. This is in contrast to our model
where, we treat the ILP class that agonizes DAF-2 as purely self-inhibiting.
The model is meant to capture the net effect of DAF-16 on all ILP
agonists, and, thus, we do not include ILP agonists that are both
activated and inhibited by DAF-16. We did however explore a model
where the dimensionless parameter $b$ is negative, implying DAF-16
has a net inhibitory effect on ILP agonists. We were able to show
that when $b$ is less than zero it is not possible to satisfy the
requirements for diffusion driven instability (see \ref{sec:AppAltMechanism}).
Given this result we propose that although INS-7 acts through a self-activating
loop, there are other self-inhibiting DAF-2 agonists that counteract
this effect.

Here, we present a specific mechanism to explain the patterns observed
in \emph{hsp-16.2} expression, however other mechanisms may contribute
to the observed patterns. We propose that serotonin at the head of
the worm leads to the release of ILPs that then diffuse through the
pseudocoelom of the worm. This is based on two key experimental findings:
(1) thermal injury leads to the release of serotonin from neurons
at the head of the worm and this release is essential for the heat
shock response in intestinal cells \citep{Tatum2015,Prahlad2009};
(2) the DAF-2 insulin like receptor is key for the heat shock response
to be activated in intestinal cells \citep{Liang2006}. Thus, there
must be a link between serotonin and the insulin-like signaling pathway.
In an alternate mechanism, serotonin may stimulate other neurons and
lead to a signal propagation through the peripheral nervous system.
However, the intestine in \emph{C. elegans} is not directly innervated
\citep{Altun2009}, so a diffusive process is still essential. It
is also possible that serotonin itself diffuses through the pseudocoelom
and directly activates the heat shock response in intestinal cells.
However, experiments have not yet revealed the presence of serotonin
receptors on intestinal cells and this mechanism does not explain
why DAF-2 is an essential component of the process. Finally, patterned
HSP expression may be, in part, caused by coelomocytes in the pseudocoelom.
Coelomocytes are large, endocytotic cells situated in relatively constant
positions in the psuedocoelomic cavity \citep{Fares2002}. Their endocytotic
properties make it likely that they function similar to immune cells.
There are 6 coelomocytes located within \emph{C. elegans}, and one
study has suggested that they may endocytose ILPs \citep{Kao2007}.
The six coelomocytes are located in pairs at the anterior, the mid
body, and the posterior of the worm. If coelomocytes endocytose a
large amount of ILPs this would affect the ILP concentration in a
spatially dependent manner and result in cell to cell variation in
the level of DAF-2 activation. Due to these potential alternate mechanisms,
future work is needed to validate the mathematical model. For example,
the model predicts that DAF-16 nuclear localization would occur in
a patterned fashion, similar to the \emph{hsp-16.2} transcription
patterns. Thus, an experiment measuring the localization of DAF-16
within individual intestinal cells would help validate the proposed
mechanism. 

Although the model provides an explanation for how patterns in protein
expression arise, the initialization of these patterns is not explicitly
included. We hypothesize that the patterning is initiated by a stimulus
at the head of the worm. The data shows \emph{hsp-16.2} transcription
is consistently higher in anterior and posterior intestinal cells
compared with middle intestinal cells. However, given random initialization
conditions, the model predicts that the reflected version of this
pattern in which expression levels are minimized at the edges would
also be observed. When performing simulations where ILPs were initially
only present at the left end of the domain, we found the resulting
pattern had an orientation that matched the data (Figure \ref{fig:simulation_c}).
This asymmetric initialization provides an explanation for why reflected
versions of the pattern are never observed in the data. Finally, we
note that as the model currently stands, ILP transcription will occur
regardless of the initialization conditions. However, under healthy
conditions this transcription should not occur. Thus, the model assumes
the stressor has affected the system in some way as to allow for ILP
transcription. Future work could involve exploring the nature of this
alteration to the system. A deeper understanding of the initialization
process could provide insight into how different magnitudes of \emph{hsp-16.2}
arise.

It is not known whether \emph{hsp-16.2} transcription directly affects
aging or if it is upstream events that are the driver of aging. In
the latter case, the model allows us to predict what the upstream
drivers may actually be. For instance, DAF-16 is a transcription factor
that regulates the expression of many proteins besides ILPs. As such,
DAF-16 nuclear translocation may play an important role in aging.
We can use the model to assess how altered parameter values might
affect the amount of DAF-16 nuclear localization. Alternatively, the
model allows us to examine how known effects of aging might alter
the system. Protein turnover dynamics are known to be affected during
the aging process \citep{Ward2002}. Thus, we can study how these
effects might propagate through the system and potentially exacerbate
the aging process. This work can be used to help understand how different
magnitudes of \emph{hsp-16.2} expression arise and why higher levels
of \emph{hsp-16.2} transcription are associated with aging. 

\section{Acknowledgments}

JMW is supported in part by an NSF GRFP and in part by the Interdisciplinary
Quantitative Biology (IQ Biology) program at the BioFrontiers Institute,
University of Colorado, Boulder. IQ Biology is generously supported
by NSF IGERT grant number 1144807. AM is supported by the National
Institute on Aging at the National Institutes of Health by grant 4R00AG045341.
The authors would also like to thank T.~E.~Johnson (University of
Colorado, Boulder) for insightful discussions and suggestions concerning
this work.

\bibliographystyle{elsarticle-harv}
\bibliography{Papers-ReactionDiffusionCElegans}

\newpage{}

\appendix
\renewcommand*{\thesection}{Appendix \Alph{section}} 

\section{Alternative Mechanism\label{sec:AppAltMechanism} }

Here we explore an alternate system in which the ILP class that acts
as a DAF-2 agonist, $B$, is repressed by DAF-16 nuclear localization.
This implies that $B$ regulates its own production through a positive
feedback loop rather than through a negative feedback loop as presented
in the main paper. The following system of equations describes this
alternative model:
\begin{align}
\frac{\partial A}{\partial t} & =F(A,B)+D_{A}\nabla^{2}A\\
\frac{\partial B}{\partial t} & =G(A,B)+D_{B}\nabla^{2}B
\end{align}
where 
\begin{align}
F(A,B)= & k_{1}-k_{2}\frac{P^{r}}{H^{r}+P^{r}}-k_{3}A\\
G(A,B)= & k_{4}+k_{5}\frac{P}{H^{r}+P^{r}}-k_{6}B
\end{align}
and 
\begin{equation}
P=\frac{V_{max}B}{k_{7}(K_{m}(1+\frac{A}{K_{i}})+B)}.
\end{equation}
The system was made dimensionless using the following substitutions
\begin{align*}
t^{*} & =D_{A}t/L^{2} & x^{*} & =x/L & d & =D_{B}/D_{A}\\
\gamma & =k_{3}L^{2}/D_{A} & u & =Ak_{3}/k_{1} & v & =Bk_{3}/k_{4}\\
p & =Pk_{7}/V_{max} & a & =k_{2}/k_{1} & b & =k_{5}/k_{4}\\
c & =k_{6}/k_{3} & h & =H(k_{7}/V_{max}) & k_{D,A} & =K_{D,A}k_{3}/k_{1}\\
k_{D,B} & =K_{D,B}k_{3}/k_{4}
\end{align*}
where all the parameters must take on positive values. This leads
to the following system of equations 
\begin{align}
\frac{\partial u}{\partial t^{*}}=\gamma f(u,v)+\nabla^{2}u\\
\frac{\partial v}{\partial t^{*}}=\gamma g(u,v)+d\nabla^{2}u
\end{align}
where 
\begin{align}
f(u,v) & =1-a\frac{p(u,v)^{r}}{h^{r}+p(u,v)^{r}}-u\\
g(u,v) & =1+b\frac{p(u,v)^{r}}{h^{r}+p(u,v)^{r}}-cv.\\
p(u,v) & =\frac{v}{k_{D,B}(1+\frac{u}{k_{D,A}})+v}
\end{align}
Ignoring diffusion and linearizing about the steady state ($u_{0},v_{0}$)
leads to the following differential equation: 
\begin{equation}
w_{t}=\gamma Aw,\quad w=\left[\begin{array}{l}
u-u_{0}\\
v-v_{0}
\end{array}\right],\quad A=\left[\begin{array}{ll}
f_{u} & f_{v}\\
g_{u} & g_{v}
\end{array}\right]_{(u_{0},v_{0})}.
\end{equation}
For diffusion driven instability to occur, the system must be stable
without diffusion, leading to the following requirements: 
\begin{align}
f_{u}+g_{v} & <0\label{eq:trace}\\
f_{u}g_{v}-g_{u}f_{v} & >0.\label{eq:det}
\end{align}
The partial derivatives of $f(u,v)$ and $g(u,v)$ are 
\begin{align}
f_{u} & =\frac{ah^{r}rk_{D,B}v^{r}}{k_{D,A}Z^{2}}Y-1\label{equ:fu1-1}\\
f_{v} & =-\frac{arv^{r-1}}{Z}+\frac{av^{r}}{Z^{2}}(rv^{r-1}+h^{r}rY)\\
g_{u} & =-\frac{bh^{r}rk_{D,B}v^{r}}{k_{D,A}Z^{2}}Y\\
g_{v} & =\frac{brv^{r-1}}{Z}-\frac{bv^{r}}{Z^{2}}(rv^{r-1}+h^{r}rY)-c\label{equ:gv1-1}
\end{align}
where 
\begin{align}
Z & =v^{r}+h^{r}(k_{D,B}(1+\frac{u}{k_{D,A}})+v)^{r}.\\
Y & =(k_{D,B}(1+\frac{u}{k_{D,A}})+v)^{r-1}\label{eq:Y}
\end{align}

Using the condition given by Equation \ref{eq:det} and a series of
algebraic manipulations, we obtain the following inequalities that
must hold for diffusion driven instability to occur
\begin{align}
f_{u}g_{v}-g_{u}f_{v}>0\implies & \left(\frac{ah^{r}rk_{D,B}v^{r}}{k_{D,A}Z^{2}}Y-1\right)\left(\frac{brv^{r-1}}{Z}-\frac{bv^{r}}{Z^{2}}(rv^{r-1}+h^{r}rY)-c\right)\nonumber \\
 & -\left(-\frac{bh^{r}rk_{D,B}v^{r}}{k_{D,A}Z^{2}}Y\right)\left(-\frac{arv^{r-1}}{Z}+\frac{av^{r}}{Z^{2}}(rv^{r-1}+h^{r}rY)\right)>0\label{eq:determinateStart-1}\\
\implies & -\frac{brv^{r-1}}{Z}+\frac{bv^{r}}{Z^{2}}(rv^{r-1}+h^{r}rY)-\left(\frac{ah^{r}rk_{D,B}v^{r}}{k_{D,A}Z^{2}}Y-1\right)c>0\\
\implies & -Zbrv^{r-1}+bv^{r}(rv^{r-1}+h^{r}rY)-\frac{ah^{r}rk_{D,B}cv^{r}}{k_{D,A}}Y+cZ^{2}>0\\
\implies & -\left(v^{r}+h^{r}\left(k_{D,B}\left(1+\frac{u}{k_{D,A}}\right)+v\right)^{r}\right)brv^{r-1}\nonumber \\
 & +bv^{r}(rv^{r-1}+h^{r}rY)-\frac{ah^{r}rk_{D,B}v^{r}c}{k_{D,A}}Y+cZ^{2}>0\\
\implies & -brv^{2r-1}-bh^{r}rv^{r-1}Y\left(k_{D,B}\left(1+\frac{u}{k_{D,A}}\right)+v\right)\\
 & +brv^{2r-1}+bh^{r}rv^{r}Y-\frac{ah^{r}rk_{D,B}cv^{r}}{k_{D,A}}Y+cZ^{2}>0\\
\implies & -bh^{r}rv^{r-1}Y\left(k_{D,B}\left(1+\frac{u}{k_{D,A}}\right)+v\right)+bh^{r}rv^{r}Y-\frac{ah^{r}rk_{D,B}cv^{r}}{k_{D,A}}Y+cZ^{2}>0\\
\implies & -bh^{r}rk_{D,B}v^{r-1}Y\left(1+\frac{u}{k_{D,A}}\right)-\frac{ah^{r}rk_{D,B}cv^{r}}{k_{D,A}}Y+cZ^{2}>0\label{eq:determinateEnd-1}
\end{align}

Furthermore, for diffusion to cause instability in the system the
following relation must hold
\begin{equation}
df_{u}+g_{v}>0.\label{equ:17052016.2}
\end{equation}
Taken together with Equation \ref{eq:trace}, this implies that $f_{u}$
and $g_{v}$ must have opposite signs. This leads to two possible
cases. In Case 1, $f_{u}>0$ and $g_{v}<0$, and in Case 2, $f_{u}<0$
and $g_{v}>0$. For Case 1, using Equations \ref{equ:fu1-1} and \ref{equ:gv1-1},
we derive the following inequalities:
\begin{align}
f_{u}>0\implies & \frac{ah^{r}rk_{D,B}v^{r}}{k_{D,A}}\left(k_{D,B}\left(1+\frac{u}{k_{D,A}}\right)+v\right)^{r-1}>Z^{2}\label{eq:case1_start}\\
g_{v}<0\implies & cZ^{2}>Zbrv^{r-1}-bv^{r}\left(rv^{r-1}+h^{r}r\left(k_{D,B}\left(1+\frac{u}{k_{D,A}}\right)+v\right)^{r-1}\right)\\
\implies & cZ^{2}>brv^{2r-1}+bh^{r}rv^{r-1}\left(k_{D,B}\left(1+\frac{u}{k_{D,A}}\right)+v\right)^{r}-brv^{2r-1}\nonumber \\
 & -bh^{r}rv^{r}\left(k_{D,B}\left(1+\frac{u}{k_{D,A}}\right)+v\right)^{r-1}\\
\implies & Z^{2}>\frac{bh^{r}rv^{r-1}}{c}\left(\left(k_{D,B}\left(1+\frac{u}{k_{D,A}}\right)+v\right)^{r}-v\left(k_{D,B}\left(1+\frac{u}{k_{D,A}}\right)+v\right)^{r-1}\right)\\
\implies & Z^{2}>\frac{bh^{r}rv^{r-1}}{c}\left(k_{D,B}\left(1+\frac{u}{k_{D,A}}\right)+v\right)^{r-1}\left(\left(k_{D,B}\left(1+\frac{u}{k_{D,A}}\right)+v\right)-v\right)\\
\implies & Z^{2}>\frac{bh^{r}rk_{D,B}v^{r-1}}{c}\left(k_{D,B}\left(1+\frac{u}{k_{D,A}}\right)+v\right)^{r-1}\left(1+\frac{u}{k_{D,A}}\right).\label{eq:case1_end}
\end{align}
In summary, for Case 1, $Z^{2}$ must satisfy the following inequality
(using Equation \ref{eq:Y}):
\begin{equation}
\frac{bh^{r}rk_{D,B}v^{r-1}}{c}\left(1+\frac{u}{k_{D,A}}\right)Y<Z^{2}<\frac{ah^{r}rk_{D,B}v^{r}}{k_{D,A}}Y.\label{eq:case1-1}
\end{equation}
Thus, for Case 1, using the fact that the determinate of $A$ must
be greater than zero (Equation \ref{eq:determinateEnd-1}) and the
upper bound on $Z^{2}$ given in Equation \ref{eq:case1-1} we have
that
\begin{align}
\frac{ah^{r}rk_{D,B}v^{r}}{k_{D,A}}Y>Z^{2}>\frac{bh^{r}rk_{D,B}v^{r-1}}{c}(1+\frac{u}{k_{D,A}})Y+\frac{ah^{r}rk_{D,B}v^{r}}{k_{D,A}}Y\\
\implies0>Z^{2}-\frac{ah^{r}rk_{D,B}v^{r}}{k_{D,A}}Y>\frac{bh^{r}rk_{D,B}v^{r-1}}{c}(1+\frac{u}{k_{D,A}})Y & .\label{eq:case1}
\end{align}
However, this inequality is not possible since all the parameter values
are greater than zero.

A similar argument holds for Case 2. Using the same steps as shown
in Equations \ref{eq:case1_start} - \ref{eq:case1_end}, but reversing
the equality sign, $Z^{2}$ must satisfy the following inequality:
\begin{equation}
\frac{ahrk_{D,B}v^{r}}{k_{D,A}}Y<Z^{2}<\frac{bhrk_{D,B}v^{r-1}}{c}\left(1+\frac{u}{k_{D,A}}\right)Y.\label{eq:case2-1}
\end{equation}
Equation \ref{eq:determinateEnd-1} and \ref{eq:case2-1} imply that
\begin{align}
\frac{bhrk_{D,B}v^{r-1}}{c}\left(1+\frac{u}{k_{D,A}}\right)Y>Z^{2}>\frac{bhrk_{D,B}v^{r-1}}{c}\left(1+\frac{u}{k_{D,A}}\right)Y+\frac{ahrk_{D,B}v^{r}}{k_{D,A}}Y\\
\implies0>Z^{2}-\frac{bhrk_{D,B}v^{r-1}}{c}\left(1+\frac{u}{k_{D,A}}\right)Y>\frac{ahrk_{D,B}v^{r}}{k_{D,A}}Y & .\label{eq:case2}
\end{align}
Since the parameters are positive, the final term in Equation \ref{eq:case2}
can not be less than zero. Thus, for this system it is not possible
for the determinate of $A$ to be greater than zero and for $f_{u}$
and $g_{v}$ to have opposite signs. Therefore, diffusion driven instability
cannot occur. 
\end{document}